\documentclass[trackchanges]{aastex7}
\usepackage{graphicx} 

\usepackage[utf8]{inputenc}
\usepackage{amsmath}
\usepackage{amssymb}
\usepackage{graphicx}
\usepackage{bm}
\usepackage{float}
\usepackage{hyperref}
\hypersetup{colorlinks=true,linkcolor=blue,filecolor=blue,urlcolor=blue,citecolor=blue}
\usepackage{orcidlink}

\begin{document}

\title{Direct Spectroscopy of 51 Eridani b with JWST NIRSpec}
\author{Alexander Madurowicz \orcidlink{0000-0001-7443-6550}}
\affiliation{University of California, Santa Cruz}
\affiliation{University of California, San Diego}
\affiliation{Space Telescope Science Institute}
\email{amadurowicz@stsci.edu}

\author{Jean-Baptiste Ruffio \orcidlink{0000-0003-2233-4821}}
\affiliation{University of California, San Diego}
\email{}

\author{Bruce Macintosh \orcidlink{0000-0003-1212-7538}}
\affiliation{University of California, Santa Cruz}
\email{}

\author{Marshall Perrin \orcidlink{0000-0002-3191-8151}}
\affiliation{Space Telescope Science Institute}
\email{}

\author{Quinn M. Konopacky \orcidlink{0000-0002-9936-6285}}
\affiliation{University of California, San Diego}
\email{}

\author{Aneesh Baburaj \orcidlink{0000-0003-3708-241X}}
\affiliation{University of California, San Diego}
\email{}

\author{Kielan Hoch \orcidlink{0000-0002-9803-8255}}
\affiliation{Space Telescope Science Institute}
\email{}

\begin{abstract}
We present high-contrast direct spectroscopy of the low-mass, cool exoplanet 51 Eridani b (2-4 M$_\textrm{Jup}$, $\sim$750 K) using JWST / NIRSpec in a fixed-slit configuration (F290LP / G395H, $3-5\,\mu$m, R$\sim$2,700). A cross correlation analysis between the continuum-subtracted data and atmospheric forward models indicates a detection of molecular signals of planetary origin at $4.8\sigma$ at the expected position and velocity of the planet. The detection of the planetary signal is driven primarily by molecular features from methane and carbon monoxide, providing the first direct confirmation of these two molecules coexisting in chemical disequilibrium in the atmosphere of 51 Eridani b. A new comprehensive atmospheric model analysis shows consistency between the ground-based IFU spectroscopy and the NIRSpec data, with the best-fit model parameters: $T_\mathrm{eff}$ = 800$^{+21.5}_{-55.5}$ K, $\log g$ = 3.75$^{+0.09}_{-0.37}$, $[\mathrm{M}/\mathrm{H}]$ = 0.7$^{+0.07}_{-0.21}$, $\textrm{C}/\textrm{O}$ = 0.458$^{+0.08}_{-0.09}$, $\log K_\mathrm{zz}$ = 3$^{+0.47}_{-0.73}$, $R_\mathrm{P}$ = 1.36$^{+0.07}_{-0.03}$ $R_\mathrm{Jup}$, $f_\mathrm{hole}$ = 0.3$^{+0.10}_{-0.07}$, and the NIRSpec errorbar inflation parameter: $\hat{e}$ = 1.74$^{+0.02}_{-0.03}$. We conclude with a discussion on the lessons learned between the fixed slit and IFU-based high contrast spectroscopic methods from our observing program, including some possibilities to improve the analysis method.
\end{abstract}


\section{Introduction}
Direct imaging is a powerful technique for detection and characterization of exoplanetary systems, providing access to wide orbital separations and enabling astrometric and spectroscopic measurements to constrain planetary bulk, orbital, and compositional properties. This ensemble of information is useful for illuminating planetary demographics \citep{Nielsen_2019} as well as distinguishing between possible formation pathways \citep{Hoch_2023}. Early simulations of JWST operating in a standard coronagraphic mode with NIRCam \citep{Carter2020} set expectations for increases in direct imaging sensitivity to very low masses ($\sim$ Saturn mass) at very wide separations ($>$50 AU). This increase in sensitivity has since been confirmed by cycle 1 survey programs \citep{bogat2025}, although candidate objects remain unconfirmed. Alternatively, NIRSpec's design was flexible for use in fixed-slit or IFU-based modes with a low-resolution prism (R = 30-330) or higher-resolution gratings (R = 500-1340 or R = 1320-3600) from 0.6-5.3 $\mu$m \citep{Jakobsen2022,Boker2022}. While NIRSpec was originally intended to image faint galaxies, others quickly realized that NIRSpec was capable of characterizing exoplanets with transmission and eclipse spectroscopy \citep{Birkmann2022}.

In this paper, we use JWST NIRSpec in an unconventional observing mode, using fixed-slit spectroscopy to image and characterize 51 Eridani b at high contrast. We build on the methods laid out in \citet{Ruffio_2024,Hoch_2024} which demonstrated spectroscopic detection and atmospheric characterization of HD 19467 B at moderate contrast with the NIRSpec IFU. This method relies on using moderate spectral resolution observations combined with joint modeling techniques to separate the stellar and planetary contributions to the total observed signal. When compared to low resolution (R$\sim$100) spectroscopy, which can detect broadband molecular absorption features from many groups of transitions at proximal wavelengths/energies, moderate resolution (R$\gtrsim$1000) spectroscopy can resolve individual molecular bands from specific rotational and vibrational transitions. When combined with high-pass filtering to remove the stellar and planetary continuum, these molecular features remain in the continuum-subtracted data as oscillations in the residuals with unique spectral fingerprints. Cross-correlation analysis between model templates with known molecular features and these residuals can reveal the existence of the planetary contribution to the total flux. This technique is capable of detecting planets at high contrast even in the absence of a coronagraph to block the starlight. This is possible because spectral correlations can be used to draw planetary signal out of the spatially variable ``speckle-noise" from scattered starlight, which is the dominant noise source for standard broadband coronagraphic imaging. The significant heritage of applying this technique on ground based observatories, including with KPIC \citep{Delorme2021, Wang_2021}, OSIRIS \citep{Ruffio_2021}, and CRIRES \citep{Snellen_2014}, led to the rapid application of this powerful approach to a space-based platform. 

The observational strategy using the fixed slits was originally chosen to address concerns that the very bright (V=5.2 mag) \citep{Simon2011} host star would saturate the detector in full frame exposures. This would lead to unrecoverable pixels at the position of the planet as ``charge bleeding" effects in the detector cause photoelectrons to spill over into neighboring pixels. By observing with the slit, not only would the starlight from the central PSF core of the host star be mitigated, but also it would enable subarray readout mode with faster cadences. This strategic observing choice has other consequences which we discuss in greater detail later, including additional exposures/pointings required for the ``speckle-side" dataset and loss of representative spatial sampling of the ``noise annulus" for estimation of the covariance of the errorbars.

The target we observe, 51 Eridani b, has been studied previously from the ground using GPI, SPHERE, and Keck/NIRC2. \citep{Macintosh2015, deRosa2015, Samland2017, Whiteford2023, BrownSevilla2023} The discovery paper \citep{Macintosh2015} reported a detection of methane absorption around the J band (1.1-1.4 $\mu$m) which made it unique among directly imaged objects around the same age and temperature, hinting at the presence of disequilibrium chemistry. While obtaining a dynamical mass measurement of 51 Eridani b from a combination of astrometry and Gaia-Hipparcos acceleration is not yet possible (De Rosa, priv. comm.), evolutionary models can estimate the planetary mass from its age and luminosity.  These models indicate that 51 Eridani b is between 2-12 M$_\textrm{Jup}$, which is one of the lowest mass directly imaged objects known. This makes 51 Eridani b ``a bridge between wider-orbit, hotter, more massive planets and planets at Jupiter-like scales" \citep{Macintosh2015}.

Previous atmospheric analyses using forward modeling found that a ``partly-cloudy" atmosphere \citep{Rajan_2017} was needed to explain the spectral energy distribution. However, retrieval analyses \citep{BrownSevilla2023, Whiteford2023} preferred clear models. The authors attribute this to the flexibility of the retrieval framework to ``mimic" cloud-like effects in the thermal structure of atmosphere. However, all of these previous analyses had assumed equilibrium chemistry or free chemistry for the retrieval. Our previous work uniquely included modeling of effects from disequilibrium chemistry and clouds together \citep{Madurowicz_2023}, in particular the disequilibrium between carbon monoxide and methane. These effects when combined can resolve the large discrepancy between the models and data for the photometric point from Keck at 4.5 micron in the Ms band. The possibility to directly detect carbon monoxide absorption features motivates the 3-5 $\mu$m spectroscopy with JWST / NIRSpec we report in this paper. The partly-cloudy disequilibrium models are consistent with the reported non-detection of 51 Eridani b with cycle 1 JWST NIRCam observations in filters F430M and F460M \citep{Balmer_2025}. 

The paper is organized as follows. In Section 2 we describe the observations that will be analyzed throughout the paper, in particular the NIRSpec spectra and accompanying ground-based data from GPI and Keck. In Section 3 we describe the methods used to reduce the new JWST / NIRSpec data including the JWST pipeline and the custom modeling package BREADS \citep{Agrawal_2023}. Section 4 contains the bulk of the analysis of the NIRSpec data products, including computations of the cross-correlation functions, ``leave one molecule out" tests to search for the presence of individual molecules such as carbon monoxide and methane, as well as a comprehensive spectral analysis of the ground and space data together in a joint likelihood framework. Section 5 concludes the paper with a discussion that summarizes our main results and their future implications.

\section{Description of Observations}
The new observations that we present in this paper were executed during JWST Cycle 2 as a part of  \href{https://www.stsci.edu/jwst/phase2-public/3522.pdf}{program 3522}. The instrument is configured to observe both a ``planet-side" and ``speckle-side" pointing for both of the 200 mas wide fixed slits S200A1 and S200A2. Since the host star is too bright for target acquisition, we use the target Gaia DR3 3205095194041180160 for wide angle target acquisition and offset from that position. The filter wheel is set to F290LP, which covers $\lambda \in [2.87, 5.27]$ $\mu$m. The disperser of choice is the high resolution (R $\sim$ 2700) grating G395H.  This enables a complete spectral reconstruction across the gap between the two NIRSpec detectors (NRS1 and NRS2), which subtends $\lambda \in [3.69, 3.79]$ $\mu$m in the ``A1" slit and $\lambda \in [3.81, 3.92]$ $\mu$m in the ``A2" slit. We name these sequences of observations in shorthand as ``a1p," ``a1s," ``a2p," and ``a2s" to concisely refer to specific combinations of slits (a1 and a2) and pointings (planet-side and speckle-side).

The detectors are configured to readout in the fastest possible configuration NRSRAPID, using the subarray modes SUBS200A1 and SUBS200A2 to avoid saturating the detector due the extremely bright nature of the target. The observation sequences are dithered with the maximum possible number of dithers, five primary positions and four additional subpixel dithers at each primary position. This dithering improves the spatial sampling of the observations, with a total of 20 dithers per sequence. Each exposure is programmed to use four groups per integration and 40 integrations per exposure for a total exposure time of $\sim$3.43 hours across all configurations.

In addition to the newly analyzed slit spectra, we also analyze previously published spectra and photometry of 51 Eridani b in a new joint analysis. The previously published results include the original J and H band GPI spectra from the discovery paper \citep{Macintosh2015}, additional GPI K1 and K2 spectra as well as Keck/NIRC2 Lp and Ms photometry that was published in \citep{Rajan_2017}, and JWST/NIRCam F410M, F430M, and F460M photometry that was published in \citet{Balmer_2025}.

\section{Data Reduction Methodology}
\subsection{JWST Pipeline}
The data from program 3522 are reduced using a combination of the JWST pipeline \citep{jwst_pipeline} and BREADS (the Broad Repository for Exoplanet Analysis, Discovery, and Spectroscopy) \citep{Agrawal_2023,Ruffio_2024}, with minor modifications and customizations specific to this program. First, the raw data or \texttt{uncal} files are downloaded using the helper tool \href{https://github.com/spacetelescope/jwst_mast_query}{\texttt{jwst\char`_mast\char`_query}}. Next, we use version 1.17.1 of the JWST pipeline to run stages 1 and 2. Stage 1 of the pipeline includes the processing steps \texttt{group\char`_scale}, \texttt{dq\char`_init}, \texttt{saturation}, \texttt{superbias}, \texttt{refpix}, \texttt{linearity}, \texttt{dark\char`_current}, \texttt{charge\char`_migration}, \texttt{jump}, \texttt{clean\char`_flicker\char`_noise}, \texttt{ramp\char`_fitting}, and \texttt{gain\char`_scale}. All of these steps are run with the default parameters with the exception of \texttt{saturation}, where we pass \texttt{n\char`_pix\char`_grow\char`_sat} = 0 instead of the default of 1. This has the effect of limiting the saturation checking from expanding to neighboring pixels. At the end of stage 1 the pipeline saves the 2 dimensional \texttt{rate} files in units of data numbers per second or DN/s.

There is a consensus in the transit spectroscopy community about the importance of an additional data processing step which we will refer to as the stage 1.5 correction since it occurs between stages 1 and 2. The idea is to correct correlated read noise or ``1/f" noise by masking the spectroscopic trace in the \texttt{rate} files and performing a column-independent linear regression on the dark or unilluminated pixels to flatten or zero out their value  \citep{Sarkar2024,Espinoza_2023,Rustamkulov_2022,Schlawin_2020,Alderson_2024,schmidt2025,Alderson_2025,Luque2024,Alam_2025,Scarsdale_2024,Gressier_2024,sikora2024,Wallack_2024,benneke2024}. We have experimented with including or not including this additional customization and found that it did not provide an improvement in the quality of our results.  We have therefore opted to forgo this step in our final analysis, electing to run the default nsclean step \citep{Rauscher_2024} instead. This may be explained in part by the continuously evolving nature of the JWST pipeline software. For reference, our results are using the 1.17.1 version of the pipeline which was just released in January 2025, after the publication of most of the works advocating for stage 1.5.

Next is stage 2 of the JWST pipeline, which includes the data processing steps: \texttt{assing\char`_wcs}, \texttt{nsclean}, \texttt{background}, \texttt{extract\char`_2d}, \texttt{srctype}, \texttt{wavecorr}, \texttt{flat\char`_field}, \texttt{pathloss}, \texttt{photom}, \texttt{pixel\char`_replace}, \texttt{resample\char`_spec}, and \texttt{extract\char`_1d}. All of these steps are run with the default parameters with the exception of \texttt{extract\char`_1d} which is skipped. The stage ends by saving the two dimensional \texttt{cal} files in units of MJy, which serve as the starting point for the continued custom post-processing we perform using BREADS.
\subsection{Baking BREADS}
The original demonstration of using BREADS to perform high-contrast spectroscopy with JWST data is shown in great detail in \citep{Ruffio_2024}, where HD 19467 B was imaged and characterized using the NIRSpec IFU. Our analysis follows very similar principles with a few minor modifications that are specific to our observations, which use the NIRSpec fixed slits instead of the IFU.  The major concept behind the BREADS framework is to model the data directly in the detector space, where we refer to the four-dimensional object $(\alpha, \delta, \lambda, F)$ as the point-cloud, where $\alpha$ is the right ascension, $\delta$ is the declination, $\lambda$ is the wavelength, and $F$ is the stage 2 calibrated flux in units of MJy. Each of these objects is implicitly a function of the detector pixel coordinates $x$ and $y$. The pixel coordinates are not to be confused with the IFU-aligned coordinate system IFU $X$ and IFU $Y$, which is simply the sky coordinates $(\alpha, \delta)$ rotated by the parameter \texttt{east\char`_2V2\char`_deg}, such that the orientation of the slit is vertical along the IFU $Y$ coordinate axis. The parameter \texttt{east\char`_2V2\char`_deg} is derived by taking the difference between the two JWST header objects \texttt{V3I\char`_YANG} $-$ \texttt{ROLL\char`_REF}. 

By modeling the data in the detector space with a combination of a spline-based continuum and empirical model for high-frequency components of the stellar spectra, we can accurately subtract the stellar contribution to the total flux at the cost of losing the continuum flux from the planet. The continuum-subtracted spectra are then cross-correlated with a template of the planetary spectrum, which contains unique signatures due to molecules such as CH$_4$, CO$_2$, and CO.  
At an effective stellar temperature around 7500 K \citep{Elliott_2024}, these molecules do not exist in the atmosphere of 51 Eridani A and reveal the existence of the planetary contribution to the flux.

\begin{figure}[H]
    \centering
    \includegraphics[width=1\linewidth]{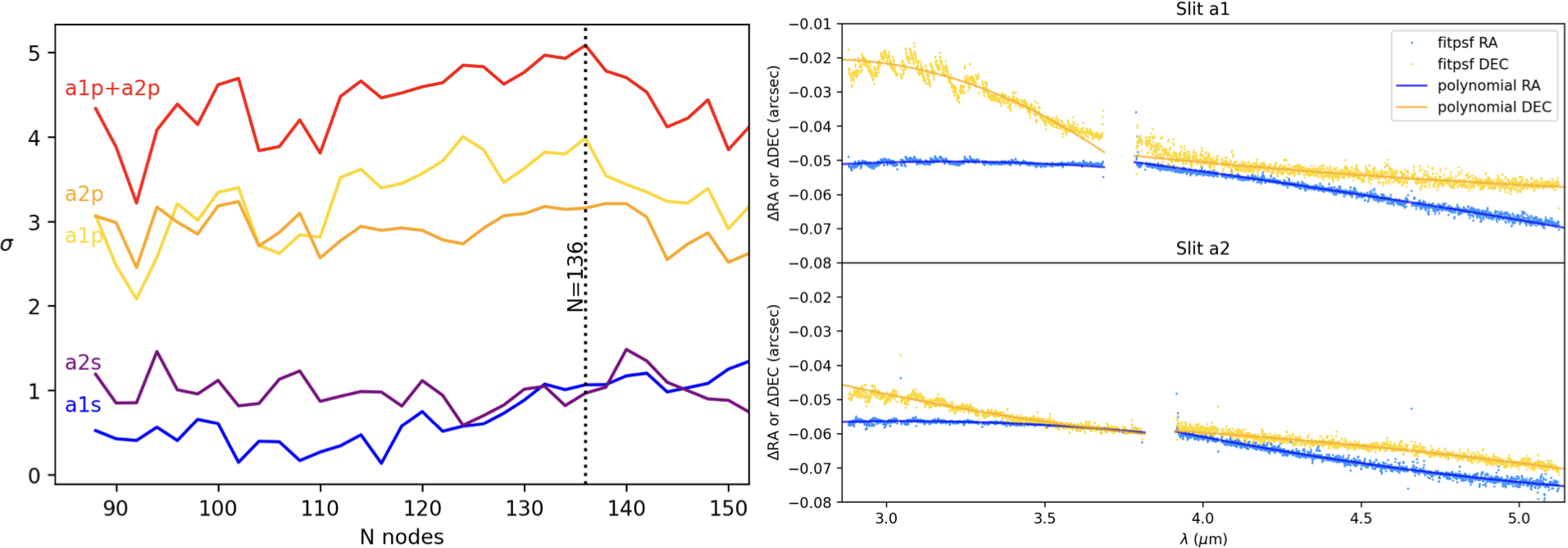}
    \caption{Diagnostic plots generated during data reduction. (Left) Detection significance as a function of the node spacing hyperparameter for each individual data sequence showcasing the robustness of the detection, as well as the optimal node spacing. (Right) Best fit coordinates in slit A1 (top) and slit A2 (bottom) and corresponding regularized polynomial fit as a function of wavelength computed during the centroiding / coordinate correction step.}
    \label{fig:reduction}
\end{figure}

One of the major free parameters in this analysis is the choice of node spacing for the spline continuum model. Due to the complexity of optimizing the reduction pipeline over this hyperparameter, we opt to only investigate uniform node spacings. This means the node spacing can be characterized by a single integer $N$ which is the total number of nodes needed to cover the entire wavelength range $\lambda \in [2.86, 5.13]$. We investigated a range of values for $N$ from 88 to 152 in increments of 2 and found that our results are somewhat insensitive to this choice, indicating the detections are robust to the choice of hyperparameters. However, the best results were obtained for N = 136, and therefore we chose this value for our final analysis. This value of N results in 58 nodes covering the wavelength ranges in NRS1 and 81 nodes covering the wavelengths in NRS2, corresponding to a wavelength spacing of $\Delta\lambda = 0.0167$ $\mu$m$/$node. For comparison, the typical length scale of a molecular feature is 0.00278 $\pm$ 0.00199 $\mu$m which is around six times narrower than the node spacing. A summary plot of the detection significance as a function of this reduction hyperparameter is shown in the left panel of Figure (\ref{fig:reduction}).

To describe the specific analysis routines in greater detail, the first step in the BREADS chain is pre-processing the data to improve the bad pixel identification. Each row of the data is high pass filtered using a median filter with a window size of 50 pixels. Then, we compute the median absolute deviation and flag all pixels greater than 50 times that value as bad pixels. The next step is to compute the sky coordinate arrays over the detector pixel indices using the \texttt{calfile.slits[0].meta.wcs} objects populated by the JWST pipeline. This step is one of the customizations needed for fixed slit pipeline. After this pre-processing, the data are selectively masked with NaNs to block out large spectral features due to hydrogen absorption in the stellar spectrum. We find the hydrogen masking improves the final quality of the detection as it was difficult to precisely estimate the exact depth of the hydrogen absorption features from the empirical stellar spectra. We speculate that this is because of the different illumination profiles of slit between the speckle and planet pointings. The exact masking we apply corresponds to central wavelengths $\lambda \in \{3.0395, 3.297, 3.741, 3.8125, 4.0525, 4.654\}$ $\mu$m with a half-width of $\delta\lambda$ = 0.005 $\mu$m. These wavelengths correspond to atomic transitions in the hydrogen atom between the states ($n_1$, $n_2$) corresponding to (5, 10), (5, 9), (5, 8), (6, 16), (4, 5), and (5, 7) respectively, which is mostly the Pfund series of absorption lines.

The next steps are to compute the continuum-normalized stellar spectra, fit and subtract the starlight continuum spline model, and interpolate the residuals onto a regular wavelength grid. Each slit/detector pairing (e.g. a1n1, a1n2, a2n1, and a2n2) is analyzed independently. One unique modification specific to the fixed slit analysis pipeline is that we compute the stellar spectra only from the speckle-side pointing dataset. If a dataset is from the planet-side sequence, we replace the typical \texttt{dataobj.star\char`_func} with the equivalent stellar spectra from a speckle-side sequence before computing the starlight subtraction. Another modification specific to the fixed slit pipeline is that we prepare two different kinds of ``combined" data objects containing all of the data from the entire observation sequence. One class of combined data objects contains the star-subtracted data from the entire sequence of observations in the planet-side and speckle-side pointings individually. There are a total of eight combined data objects in this set, two for each slit, detector, and pointing. These are used to extract the spectrum of the planet and the spectrum of the noise in the speckle pointing since the starlight continuum has been removed. The other combined data objects contain the continuum-preserving data from both the planet and speckle pointings combined together. These combined data objects are used to apply a coordinate correction / centroiding operation to more precisely calibrate the point cloud coordinates and rely on the starlight continuum remaining to precisely fit for that centroid.

The centroiding / coordinate correction uses the continuum-preserving, planet$+$speckle combined sequences and runs the \texttt{fitpsf} module to fit a synthetic WebbPSF to the point cloud as a function of $\lambda$ and find the best-fit coordinate correction $(\Delta\alpha, \Delta\delta)$. The values range from -0.02 to -0.08 arcsec in both right ascension and declination with a small amount of curvature along the spectral axis. The empirical best-fit coordinate correction as a function of wavelength is regularized by fitting a second-order polynomial to smooth out systematic errors caused by spatial undersampling. The coordinate correction as a function of wavelength is shown in the right plot of Figure (\ref{fig:reduction}). It is important to recognize that this coordinate correction is sub-pixel and so represents a relatively small calibration correction to the existing JWST WCS system, but necessary in order to precisely align the expected position of the planet in the center of the sub-pixel dithering pattern, which has offsets of around $\pm$0.025 arcsec in both directions. The planet's positional uncertainty based on orbit fitting is around .0002 arcsec \citep{whereistheplanet}, which is an order of magnitude more precise than the WCS coordinate calibration, since the typical JWST pointing accuracy is a ``few mas" (i.e. 0.002-0.003") after target acquisition \citep{Rigby2023}.

After the coordinate correction is applied, the final step in the BREADS chain is to perform a spectral extraction. The \texttt{fitpsf} module is used to find the best fit flux (as a function of position and wavelength) for the starlight-subtracted residual data, using a synthetic WebbPSF. The position locations to perform the fit are an input parameter and it is important to have previously performed the regular wavelength interpolation so that \texttt{fitpsf} can treat each columnar slice of the combined data objects as an independent wavelength channel. This regular wavelength grid interpolation is known to result in a 2 pixel wide covariance in the final extracted errors, which was demonstrated in the appendix of \citep{Ruffio_2024}. A major disadvantage of using the slit compared with the IFU is that there are substantially less spatial locations with similar flux that are sampled to reliably estimate the empirical error covariance. The final result of the BREADS chain are the spectral extraction products which are the continuum-subtracted residual fluxes as a function of position and wavelength and their corresponding uncertainties. 
\section{Spectral Analysis and Model fitting}
\subsection{Cross Correlation Analysis}
In order to analyze the extracted spectra, we rely on a template-based cross-correlation technique. We employ substellar atmosphere models from Sonora Elf Owl \citep{Mukherjee_2024,mukherjee_2023}. We use a combination of the \texttt{species} software package \citep{Stolker2020} as a backend and the \texttt{breads.atm\char`_utils.miniRGI} module for convenience to load and gaussian kernel broaden the model spectra at a resolution of R = 2700. Because the extracted spectra have been effectively high-pass filtered by the spline-based continuum subtraction, it is important that the models are equivalently high-pass filtered with the same spline node configuration. It is possible to accomplish this with the BREADS utility function \texttt{breads.utils.filter\char`_spec\char`_with\char`_spline}.
Using a model template with the following parameters: $T_\mathrm{eff} = 900$ K, $\log \big( \frac{g}{\mathrm{cm}/\mathrm{s}^2} \big) = $ 3.5, $[\mathrm{M}/\mathrm{H}] = 0.5$, $\textrm{C}/\textrm{O} = 0.458$, and $\log \big(\frac{K_\mathrm{zz}}{\mathrm{cm}^2\mathrm{s}}\big) = 2$, we compute the cross correlation function with the following equation:
\begin{equation}
    \mathrm{CCF}\ (\mathrm{IFU}\ Y, \mathrm{RV}) = \frac{\sum_\lambda \frac{d_\lambda m_\lambda}{e_\lambda^2}}{\sqrt{\sum_\lambda \frac{m_\lambda^2}{e_\lambda^2}}}
\end{equation}
where $d$ represents the data, $e$ is the error on the data, and $m$ represents the model. The data and its error as a function of IFU $Y$ are directly taken from the spectral extraction products, while the model $m$ is implicitly a function of the model parameters and the planetary radial velocity. The radial velocity has an effect which is equivalent to stretching the effective wavelengths of the model under the following remapping: $ \lambda \rightarrow (1 + \frac{\mathrm{RV}}{c})\lambda$, which we accomplish by resampling / interpolating the model with the module \texttt{np.interp}. The CCF equation essentially represents fitting the template model to the data in a linear least squares fashion, and taking the ratio of the flux-rescaling constant to the uncertainty on the fit over the IFU $Y$ - RV phase space (see also Equation (5) in \cite{Ruffio_2019}). This means that it is not strictly a cross-correlation function according to the standard mathematical definition, since the standard CCF relies on an additive shift for the dependent variable but instead the RV parameter here multiplicatively stretches the model function in the wavelength domain. Regardless of this nuance, the CCF as we have defined it is a very useful diagnostic to elucidate the statistical correlations between the model and data and is often referred to as a cross-correlation function in the literature. These CCFs are computed for each dataset individually, but can then be
\begin{figure}[H]
    \centering
    \includegraphics[width=1\linewidth]{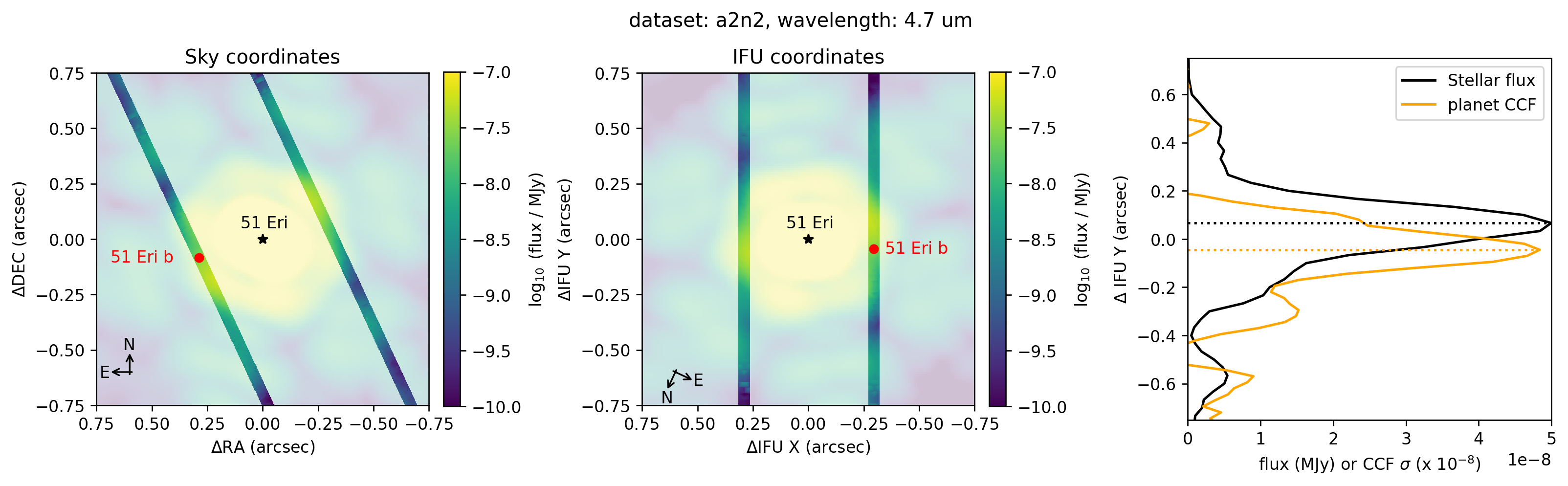}
    \includegraphics[width=1\linewidth]{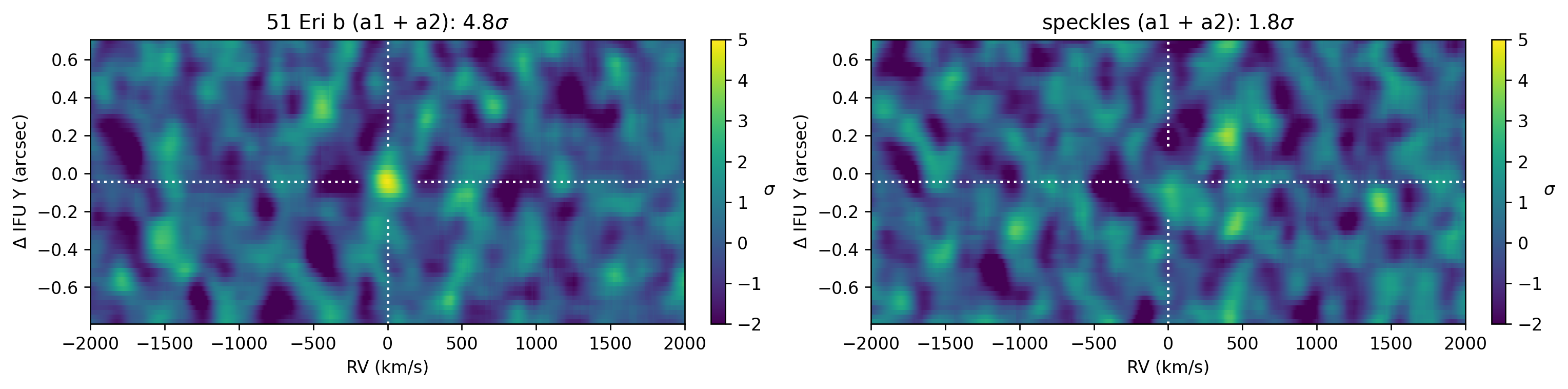}
    \caption{Detection of the planet 51 Eridani b from moderate resolution spectroscopic observations with JWST/NIRSpec. Top: Visualization of the point cloud in sky coordinates (left) and IFU-aligned coordinates (center). In the background at half opacity is the best fit synthetic WebbPSF from the coordinate correction / centroiding step. In the foreground at full opacity is an interpolation of the continuum-preserving point-cloud for the combined planet and speckle side observations. The ``slit" appears $\sim$0.05 arcsec wide in this figure because that is the scale subtended by the sub-pixel dither pattern, not the full 0.2 arcsec width of the physical slit aperture. (Right) A demonstration of the offset in IFU $Y$ between the peak intensity of the stellar PSF and the location of the peak of the cross correlation function, indicating the signal is not a result of amplified stellar noise. Bottom: Cross-correlation based detection maps in the IFU $Y$ and planetary radial velocity phase space. (Left) Joint detection map for all planet side datasets, (Right) joint detection map for all speckle side datasets. The white dashed lines indicate the expected position and velocity of the target.}
    \label{fig:pointcloud_CCF}
\end{figure}
\noindent combined using a inverse-variance weighted-mean of the best fit flux and uncertainty. In order to ensure the accuracy of the noise model for the CCF, we rescale by dividing by the standard deviation of the ``wings" of the CCF, i.e. where the $| \mathrm{RV} | > 500$ km/s. 

The CCFs for both the joint slit a1 + a2 datasets for both the planet-side and speckle-side pointings are shown in Figure (\ref{fig:pointcloud_CCF}). This figure demonstrates the clear detection of 51 Eri b in the planet-side dataset at a level of 4.8$\sigma$, localized in the IFU $Y$-RV phase space at its expected position. If the signal was caused by systematic residuals from the subtraction of the diffracted starlight, we would expect the peak of the CCF to be aligned with the peak of the starlight intensity. We verify that the two are not co-located in the upper right panel of Figure (\ref{fig:pointcloud_CCF}), which provides additional evidence that the signal is of planetary origin.
\subsection{Molecular Detection Experiments}
\begin{figure}[b]
    \centering
    \includegraphics[width=1\linewidth]{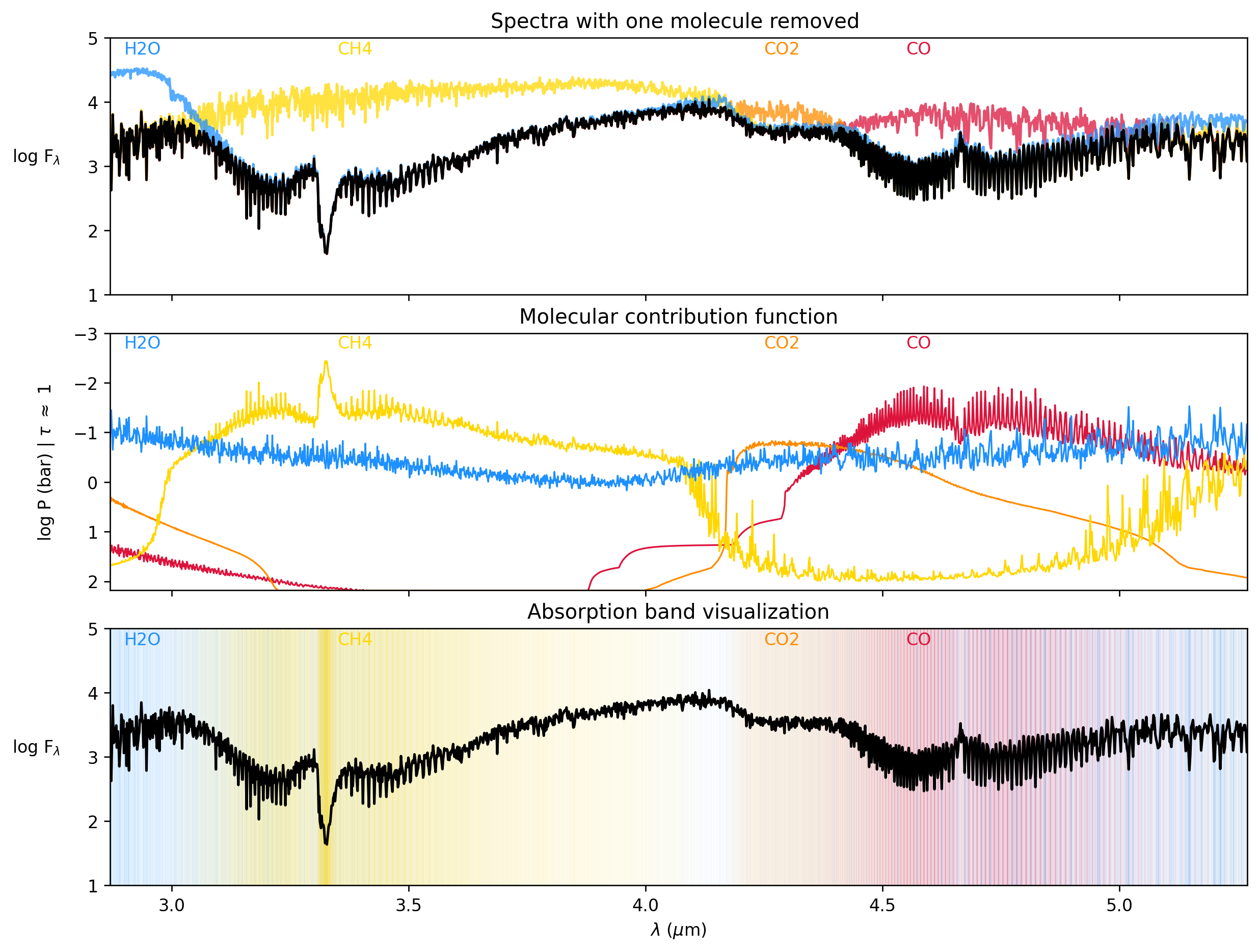}
    \caption{(Upper) Template spectra used for computing the CCF shown with perturbations from removing individual molecules in the radiative transfer calculation. (Middle) Molecular contribution functions calculated using \texttt{picaso.jdi.get\_contribution}. The curve shows the pressure layer where the optical depth per species is approximately unity. (Lower) A visualization where the absorption features are shown per species using colored banding with variable opacity to denote deeper absorption.}
    \label{fig:contribution}
\end{figure}
In addition to the spatial misalignment between the peak of the stellar flux and the peak of the planetary spectrum CCF, another line of evidence for the signal's planetary nature is to show that it originates from molecules that are expected in the planetary spectrum. At the effective temperature $\sim$750 K in the NIRSpec wavelengths, these are primarily methane (CH$_4$) and carbon monoxide (CO), although there are smaller contributions from water (H$_2$O) and carbon dioxide (CO$_2$). Other molecules which are possibly relevant at these wavelengths include hydrogen sulfide (H$_2$S) and phosphine (PH$_3$) although they are expected to be subdominant absorbers due to their lower relative abundance.

\begin{figure}[t]
    \centering
    \includegraphics[width=1\linewidth]{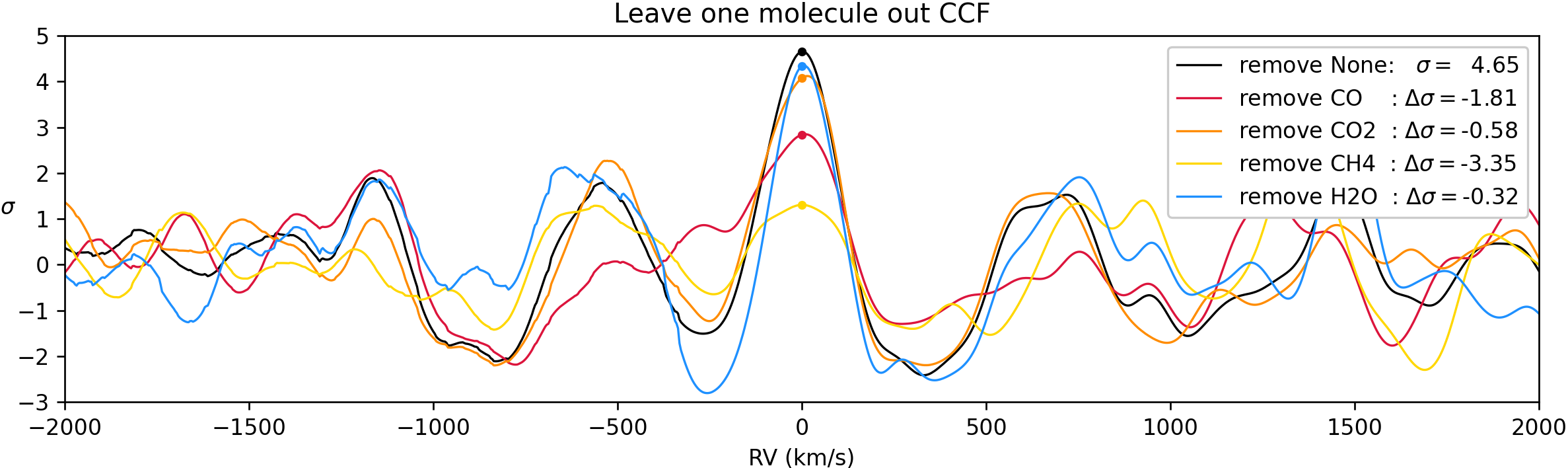}
    \includegraphics[width=1\linewidth]{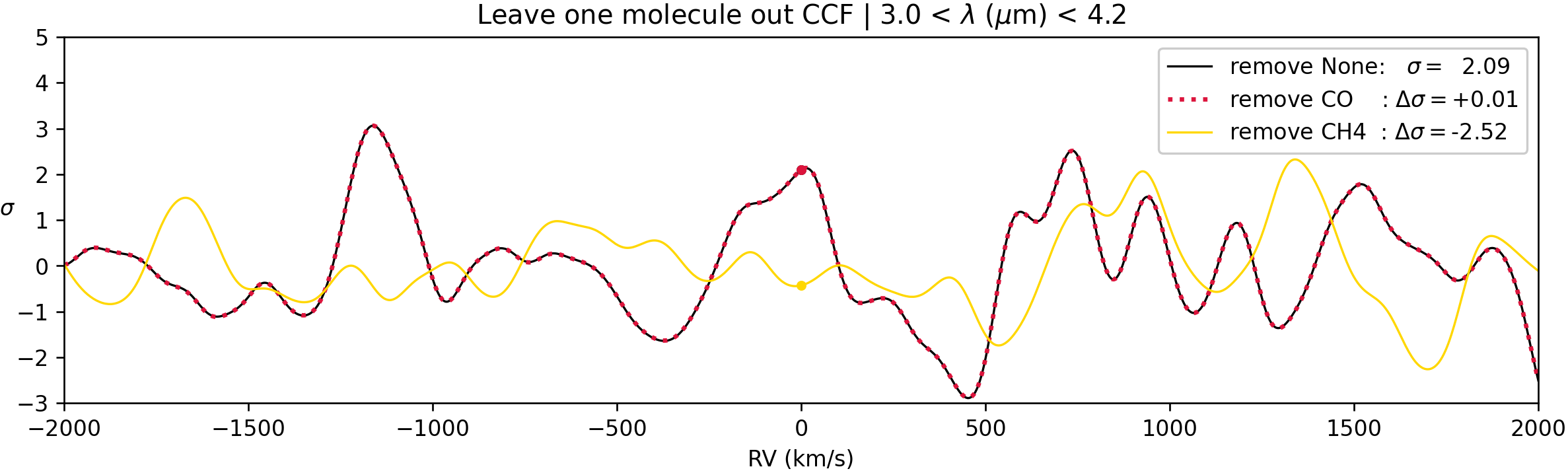}
    \includegraphics[width=1\linewidth]{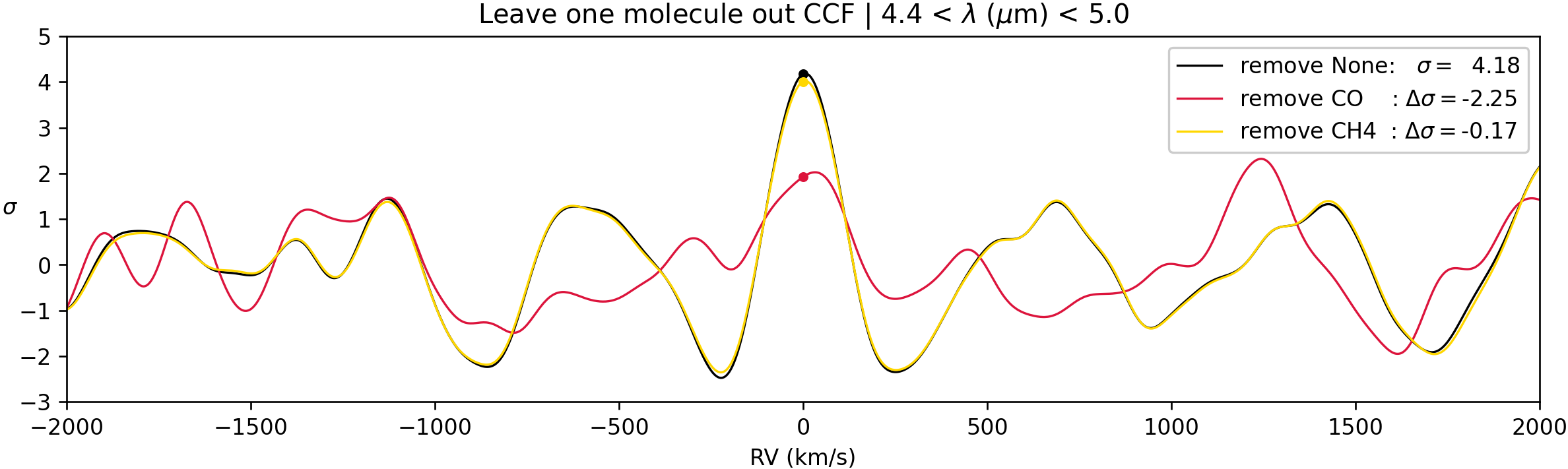}
    \caption{Searching for individual molecules in the atmosphere of 51 Eridani b with  ``leave one molecule out" cross correlations. (Top) Cross correlation functions computed on all four templates showing the effect of removing certain molecules from the atmospheric model across the entire wavelength range. (Middle) Restricted wavelength test in the region of CH4 dominance. (Bottom) Restricted wavelength test in the region of CO dominance. }
    \label{fig:CCFs}
\end{figure}

To demonstrate the effects these molecules have on the significance of the CCF detection, we perturb the spectral model of the planetary atmosphere using \texttt{PICASO} \citep{Batalha2019}. We start by loading the Sonora Elf Owl model atmosphere with parameters $T_\mathrm{eff} = 900$ K, $\log \big( \frac{g}{\mathrm{cm}/\mathrm{s}^2} \big) = $ 3.5, $[\mathrm{M}/\mathrm{H}] = 0.5$, $C / O = 0.458$, and $\log \big(\frac{K_\mathrm{zz}}{\mathrm{cm}^2\mathrm{s}}\big) = 2$. This is accomplished in \texttt{PICASO} using \texttt{jdi.xr.load\_dataset} and \texttt{jdi.input\_xarray}. It is important that we use the 2025 updated opacity tables \texttt{all\_opacities\_0.6\_6\_R60000.db} in order to accurately reproduce the published Elf Owl spectra with relative errors on the order of $\sim 10^{-5}$. These opacity tables have not been published yet but should be at some point and in the mean time can be requested by contacting the developers (Batalha, priv. comm.). For our simple test, we simply set the relative abundances for an individual molecule to zero at all pressure layers to perform our ``leave one molecule out" comparison. These spectra and additional visualizations such as the molecular contribution function are shown in Figure (\ref{fig:contribution}), while the CCFs computed using these spectra are shown in Figure (\ref{fig:CCFs}) alongside two additional tests which restrict the wavelengths to the regions of CH4 and CO dominance.
In the upper panel of Figure (\ref{fig:CCFs}) one can see how leaving out any individual molecule from the spectra produces a corresponding decrease in the resulting peak of the cross correlation function, which indicates that the features from that molecule are relevant and contribute to the detection of the planetary signal. So the detection of the planetary signal is not just from a single molecule, but from a combination of multiple relevant molecules, all of which could not exist in a stellar atmosphere. In this test, we measure the impact of an individual molecule by computing the difference in significance between the CCF detection using the full model compared to a model with that molecule missing: $\Delta \sigma = \sigma_\textrm{leave one molecule out} - \sigma_\textrm{full model}$. We find: $\Delta\sigma_\textrm{CH$_4$} = -3.35$, $\Delta\sigma_\textrm{CO} = -1.81$, $\Delta\sigma_\textrm{H$_2$O} = -0.32$, and $\Delta\sigma_\textrm{CO$_2$} = -0.58$. This test shows how the features from methane and carbon monoxide are the most important in detecting the planetary signal, and for the first time, a direct  detection of these two molecules in chemical disequilibrium in 51 Eridani b. The test also shows what we will call a ``consistent non-detection" for water and carbon dioxide, in that there is simply not enough evidence at the current level of sensitivity to claim a detection of these molecules on their own, although having them in the model does provide an improved detection. 

To make the detection significance of the individual molecules more precise, we perform additional restricted wavelength tests in the two lower panels of Figure (\ref{fig:CCFs}). The middle panel shows the CCF computed in the region $3.0 < \lambda < 4.2$ which is the range where CH4 is the dominant absorber, while the bottom panel shows the CCF in the region $4.4 < \lambda < 5.0$ where CO is the dominant absorber. In these tests one can see how removing the dominant molecule from the spectra greatly reduces the significance of the detection, and removing the irrelevant molecule does not alter the detection significantly. Furthermore, since these molecules are the primary absorber at these wavelengths, we can interpret the detection significance in this restricted wavelength test as the detection signficance of the molecules individually, implying $\sigma_\mathrm{CH4} = 2.1$ and $\sigma_\mathrm{CO} = 4.2$.

Perhaps even more interesting than just how much each molecule contributes to the total detection significance is how that is related to the total fractional bandpass where that molecule's features dominate the absorption spectra. For example, by counting the total number of wavelength channels where a molecule's contribution function (Figure (\ref{fig:contribution}) upper panel) is at a higher altitude than all of the other molecules (or equivalently, lower pressure layer) we can estimate the total fractional bandpass per molecule where that molecule's features are dominating the spectral shape. This is simply the ratio of the number of channels where its contribution function is extremal compared to the total number of channels. We find the following values for the fractional bandpass domination: $f_\textrm{CH$_4$} = 47.8\%$, $f_\textrm{CO} = 20.9\%$, $f_\textrm{H$_2$O} = 22.5\%$, and $f_\textrm{CO$_2$} = 8.8\%$. This shows how the methane and carbon monoxide features are expected to play a much larger role in our expectations of what the planetary signal should be when compared with carbon dioxide. 

The only real anomaly in this story is the water features, which have a similar fractional bandpass domination as the carbon monoxide, but are not detected at a similar level in the metric of $\Delta\sigma$. We attribute this fact to the idea that the water features are dominant at the edges of the spectral bandpass, a region of the spectra which is difficult to model accurately with the spline-continuum, due to increased curvature of the spectral trace in the detector. This implies that the systematic errors in those regions of the spectra are larger than in the central wavelengths where the spline-continuum subtraction is more accurate. If we eliminate the edges from our consideration, we then find that $f_\textrm{H$_2$O} = 4.3\%$, which puts it at a lower expected relevance than the CO$_2$, which is consistent with our findings in $\Delta\sigma$.

The review article \citep{snellen2025} discusses detection speed in the context of high resolution observations of exoplanet atmospheres for certain idealized cases. In their ``idealized molecular" case the detection speed is linearly proportional to $\lambda_\textrm{span}$, the bandwidth of the molecule under consideration. This idealized case consists of many narrow and well separated lines and they acknowledge this inexactitude by noting that certain ``wavelength regions will be richer in spectral features than others, depending on the type of planet." While this notion corroborates what we have found with a roughly linear relationship between the fractional bandpass domination for individual molecules and their corresponding impact on the CCF detection sigma we computed with $\Delta\sigma$, it is important to state this caveat that it would not necessarily be true for opacity sources with qualitatively different features. For example clouds have broadband absorption and lack many narrow features which would help identify them. Additionally the CO$_2$ and H$_2$O features appear to be qualitatively flatter when compared with CO, possibly due to an increase in the density of possible molecular state transitions in the wavelengths due to their triatomic nature, which could contribute a similar effect. This can be seen examining the molecular contribution functions in Figure (\ref{fig:contribution}).

\subsection{Comprehensive Spectral Analysis}

While the NIRSpec data is sufficient to provide a detection of 51 Eridani b on its own and investigate the molecules present in its atmosphere, it is also useful to combine this dataset with additional ground based spectroscopy and photometry to characterize the planet more comprehensively. Using all of the data described in Section 2, we compute model likelihoods over a grid of model parameters: ($T_\mathrm{eff}$, $\log g$, $[\mathrm{M}/\mathrm{H}]$, $\textrm{C}/\textrm{O}$, $\log K_\mathrm{zz}$, $R_\mathrm{P}$, $\hat{e}$, and $f_\mathrm{hole}$). The first five are intrinsic to the model grid itself, the effective temperature, surface gravity, metallicity, carbon to oxygen ratio, and vertical eddy diffusion coefficient, while $R_\mathrm{P}$ is the planet radius, $f_\mathrm{hole}$ is the cloud hole fraction, and $\hat{e}$ is a nuisance parameter which represent the errorbar inflation --- a rescaling constant which increases the size of the uncertainties of NIRSpec extracted fluxes. The cloud hole fraction represents the relative weighting between two distinct radiative transfer calculations, one with and the other without clouds \citep{Marley2010},
\begin{equation}
    F_\lambda = (1 - f_\mathrm{hole})F_{\lambda\mathrm{,cloudy}} + (f_\mathrm{hole})F_{\lambda\mathrm{,cloudfree}}
\end{equation}
where the clouds are computed using Virga \citep{batalha2025} using a sedimentation efficiency of 10. The errorbar inflation modifies the calculation of the log likelihood by introducing an additional constant
\begin{equation}
    \log \mathcal{L} \propto N_\mathrm{NIRSpec} \frac{1}{\hat{e}} - \frac{1}{2}\chi^2
    \label{eq:likelihood}
\end{equation}
\noindent where $N_\mathrm{NIRSpec}$ is the number of points in the NIRSpec data for which the errorbars are inflated. This term originates from the constant term which normalizes the traditional gaussian likelihood being modified $\frac{1}{\sqrt{(2\pi\sigma^2)}} \rightarrow \frac{1}{\sqrt{2\pi(\hat{e}\sigma)^2}}$. Here $\chi^2$ is defined using the formalism laid out previously in \citet{Madurowicz_2023}, where the covariances computed in \citet{Rajan_2017} are used to compute $\chi^2$ for the GPI IFS spectra in the J, H, K1, and K2 bands, no covariance is used for the photometry from Keck/NIRC2 or JWST/NIRCam, and a new term is added for the NIRSpec data in particular.

\begin{equation}
    \chi^2 = \sum_\textrm{GPI} v^T C^{-1} v + \sum_\textrm{phot} \Big(\frac{v}{\sigma}\Big)^2 + \chi^2_\textrm{NIRSpec}
    \label{eq:chi2}
\end{equation}
\noindent Here $v$ represents a vector of residuals between the model and data
\begin{equation}
    v = \Big(\frac{R_\textrm{planet}}{d_*}\Big)^2 F_{\lambda,\textrm{model}} - F_{\lambda,\textrm{observation}}.
\end{equation}
\noindent The NIRSpec term is slightly modified, using a covariance free estimate in the first iteration, and then a simple covariance model in the second iteration, like so:
\begin{equation}
    \chi^2_\textrm{NIRSpec} = 
    \sum_{\textrm{NIRSpec}}\begin{cases}
    \Big(\frac{\hat{v}}{\hat{e}\sigma}\Big)^2, \qquad \textrm{iteration 1} \\
    \hat{v}^T C^{-1} \hat{v}, \quad \textrm{iteration 2}
    \end{cases}
\end{equation}
This is because the slit-based observation strategy lacks adequate spatial sampling to reliably estimate the covariance model from noise spectra of empty space observed with equivalent stellar flux, how it was done previously in IFU based observations. Instead, we estimate the covariance from the autocorrelation of the data-model residuals after the first round of computing the likelihood, so the covariance is included only in the second iteration. The residual vector $\hat{v}$ is also modified to include spline-subtraction of the model, so it can be appropriately compared to the spline subtracted data.
\begin{equation}
    v = \Big(\frac{R_\textrm{planet}}{d_*}\Big)^2 \mathcal{SF}(F_{\lambda,\textrm{model}}) - F_{\lambda,\textrm{observation}},
\end{equation}
where $\mathcal{SF}$ is our representation of the spline filter \texttt{breads.filter\_spec\_with\_spline} function which removes the continuum using the spline model defined previously.

\begin{figure}[H]
    \centering
    \includegraphics[width=1\linewidth]{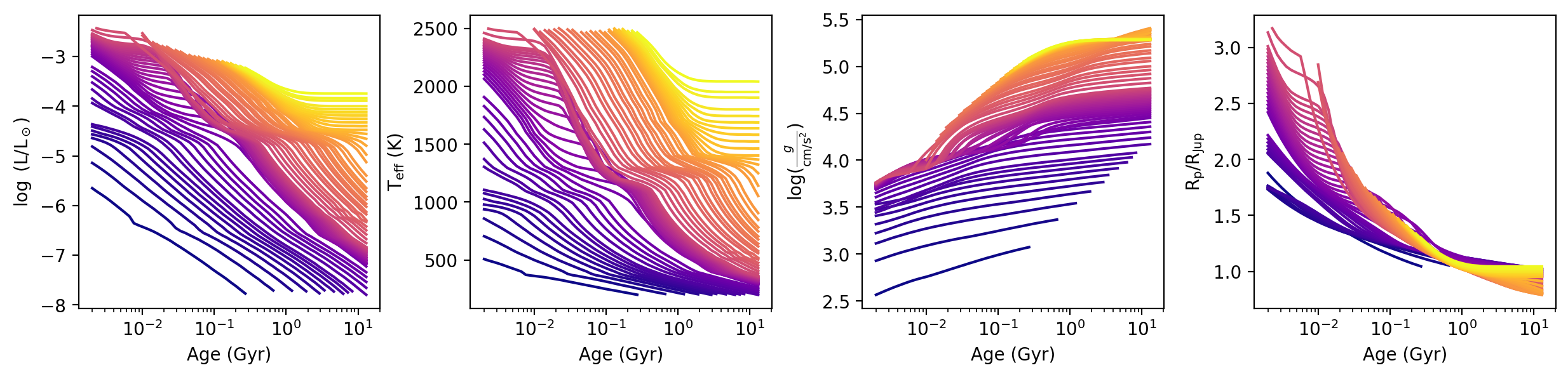}
    \includegraphics[width=1\linewidth]{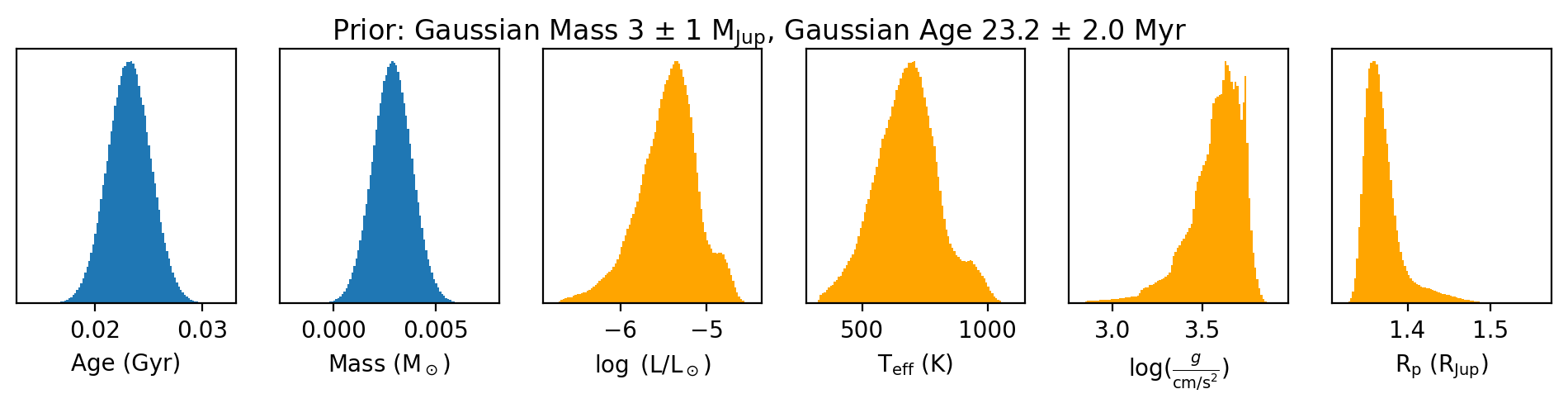}
    \caption{(Upper) Four primary parameters of the Diamondback evolutionary model grid from left to right: Luminosity, Effective Temperature, Surface gravity, and Radius as a function of age (x-axis) and mass (color). Mass ranges from 0.0005 to 0.08 M$_\odot$ going from dark blue to gold or equivalently 0.524 Mjup to 84 Mjup. (Lower) On the left, the assumption on Mass and Age for the prior distribution, and on the right the interpolation of that distribution onto the four output parameters of the distribution, which forms the prior on effective temperature, surface gravity, and radius. The prior is assumed flat over the entire grid on all of the other atmospheric parameters.}
    \label{fig:diamondback}
\end{figure}

\begin{figure}[H]
    \centering
    \includegraphics[width=1\linewidth]{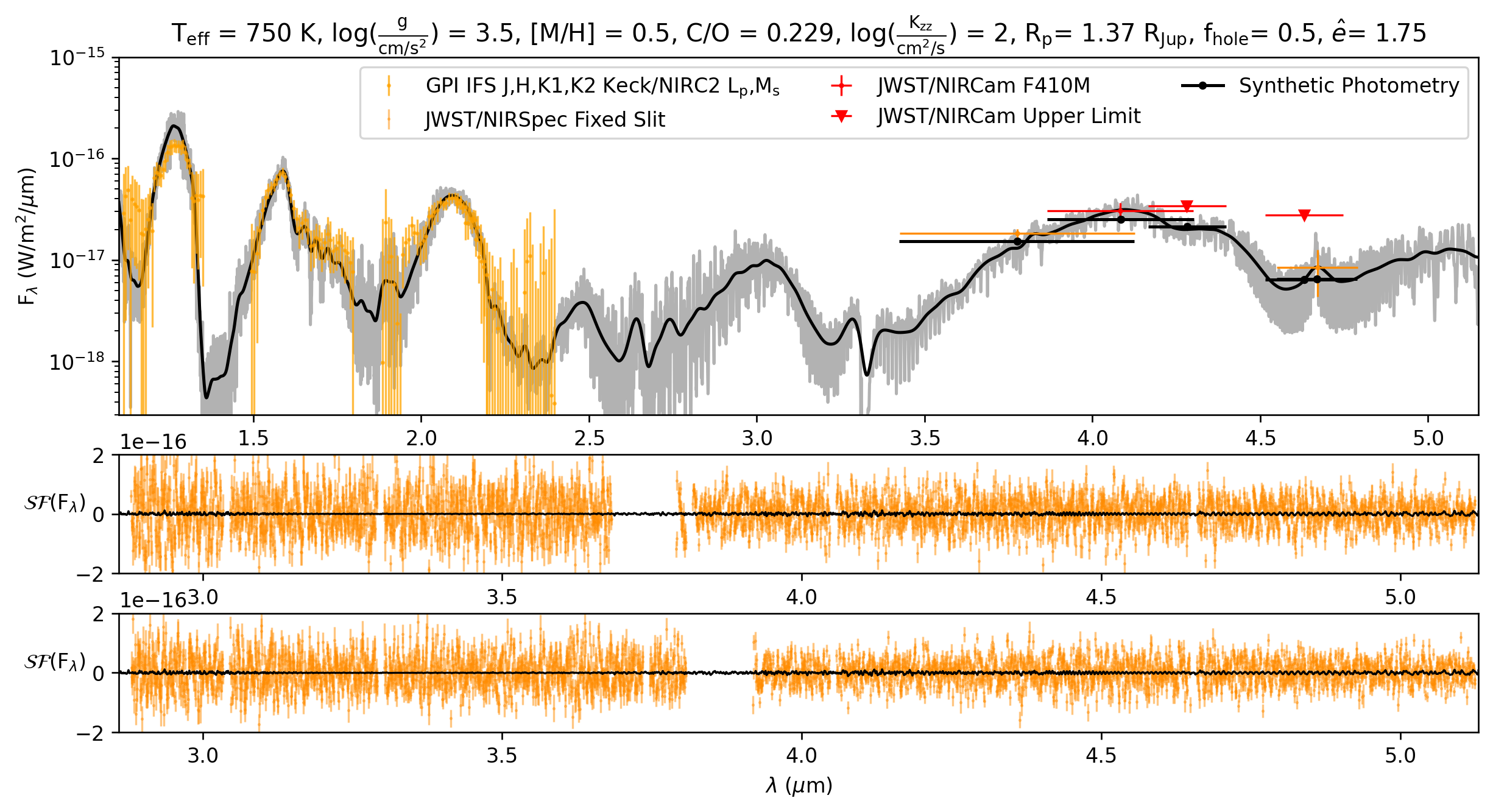}
    \caption{Best fit model spectra from the first iteration of computing likelihood with no subgrid interpolation or NIRSpec covariance shown in black at R=100 and in grey at R=2700 in the upper panel and continuum-subtracted in black at R=2700 in the two lower panels. (Upper) GPI IFS spectra in bands J,H,K1, and K2, KECK/NIRC2 Lp and Ms photometry, and JWST/NIRCam F410M, F430M, and F460M photometry in red. (Middle) NIRSpec continuum normalized spectra from slit A1 and slit A2 (Lower).}
    \label{fig:spectra}
\end{figure}

\begin{figure}[H]
    \centering
    \includegraphics[width=1\linewidth]{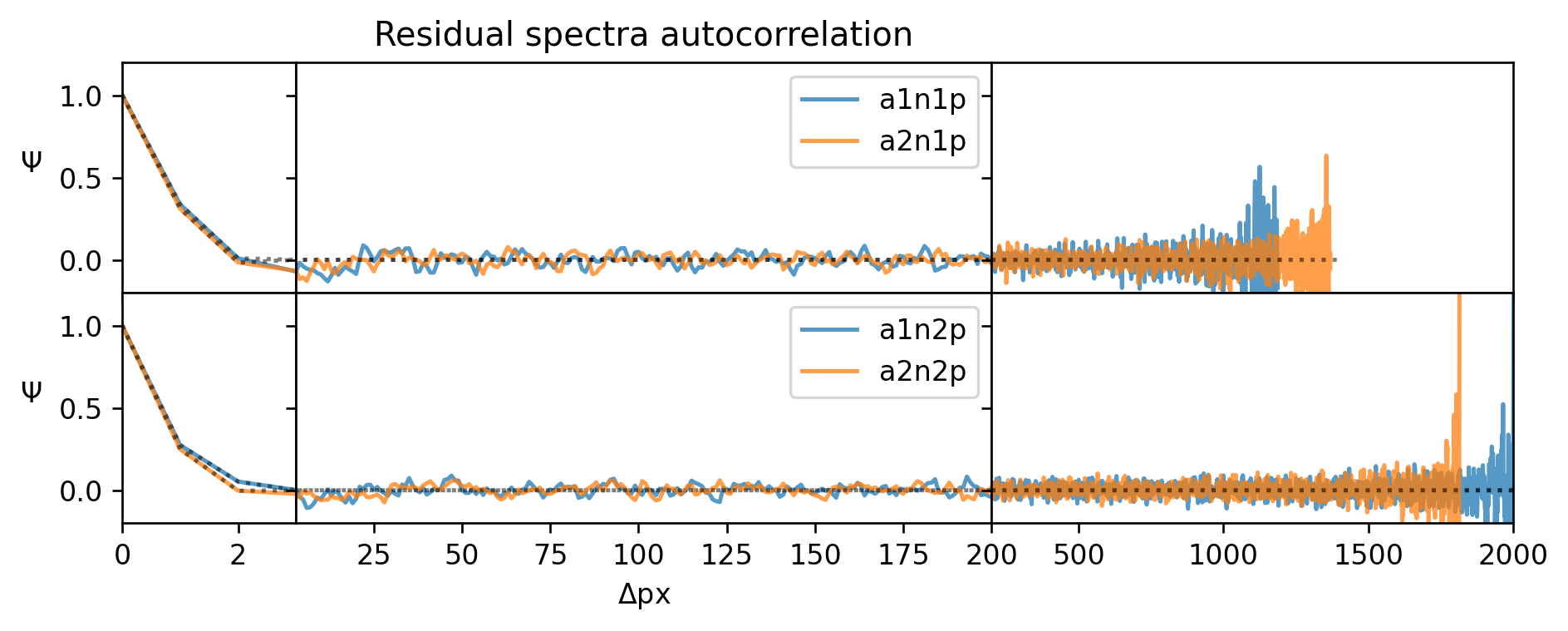}
    \caption{Correlation profile $\Psi$ for both datasets (slit a1 and a2) on detector NRS1 (upper) and NRS2 (lower) as a function of $\Delta$ pixels along the wavelength dimension.}
    \label{fig:correlation}
\end{figure}

In the first iteration of our spectral model analysis, the likelihood defined in Equation (\ref{eq:likelihood}) is calculated over the entire T-type Elf Owl model parameter space for effective temperature, surface gravity, metallicity, carbon to oxygen ratio, and vertical eddy diffusion coefficient. The exact values of these parameters appear in Table (\ref{tab:parameter_iteration}), in addition to external parameters for the planet radius, cloud hole fraction, and errorbar inflation which we define. 
\begin{table}[H]
    \centering
    \begin{tabular}{c|c|c|c}
        Description & Parameter & Iteration 1 & Iteration 2 \\
        \hline
        Effective Temperature & T$_\textrm{eff}$ $(\textrm{K})$ & [550,  600,  650,  ...  950, 1000, 1100, 1200] & [650, 675, 700, ... , 850] \\
        Surface Gravity & $\log_{10} \frac{g}{\textrm{cm}/\textrm{s}^2}$ & [3.25, 3.5, 3.75, ... 5.25, 5.5] & [3.25, 3.375, 3.5, ... , 4.0] \\
        Metallicity & $[\textrm{M}/\textrm{H}]$ & [-1, -0.5,  0,  0.5,  0.7,  1] & [0, 0.1, 0.2, ... , 1.0] \\
        Carbon to Oxygen Ratio & $\textrm{C}/\textrm{O}$ & [0.229, 0.458, 0.687, 1.145] & [0.229, 0.3435, 0.458, 0.5725, 0.687]\\
        Eddy Diffusion Coefficient & $\log_{10} \frac{K_\textrm{zz}}{\textrm{cm}^2/\textrm{s}}$ & [2, 4, 7, 8, 9] & [2, 2.5, 3, 3.5, 4, 4.5] \\
        Planet Radius & $\log_{10} \frac{R_\textrm{p}}{\textrm{R}_\textrm{Jup}}$ & [0., 0.0125, 0.025, ...,  0.2375, 0.25] & $\log_{10}$[1.26, 1.28, 1.3, ..., 1.5] \\
        NIRSpec Errorbar Inflation & $\hat{e}$ & [1.5 , 1.55, 1.6, ..., 1.95, 2] &  [1.66, 1.68, 1.7, ..., 1.8] \\
        Cloud Hole Fraction & $f_\textrm{hole}$ & [0, 0.05, 0.1, ..., 0.95, 1.] & [0.2 , 0.25, 0.3, ..., 0.7]
    \end{tabular}
    \caption{Discrete vectors defining the grid for each atmospheric model parameter during the first and second iteration of the likelihood computation.}
    \label{tab:parameter_iteration}
\end{table}
The best fit model is chosen from this selection as the model with the largest value of the posterior probability, which is the product of the likelihood from Equation (\ref{eq:likelihood}) and the prior probability, which is estimated from the Sonora Diamondback evolutionary model prior \citep{Morley_2024, morley_2024_zenodo} and appears in Figure (\ref{fig:diamondback}). We adopt the mass posterior of 2-4 $M_\mathrm{Jup}$ from \citep{BrownSevilla2023} and the age posterior of \citep{Elliott_2024} of $23.2\pm2.0$ Myr.
The best fit model is shown in Figure (\ref{fig:spectra}). A couple of things to note about this spectra: first, that it fits the ground based GPI IFS spectra as well as the mid-IR photometry rather well. Second, while it is difficult to visually verify the correlations between the NIRSpec data and spline-filtered model, which is the purpose of the cross-correlation function, the best fit errorbar inflation parameter $\hat{e}$ is only $1.75$, which means the photon noise errorbars estimated from the JWST pipeline are less than factor two off the total error. This total error includes not only photon noise but also systematic errors from imperfections in the spline filter subtraction. Thirdly, the best fit radius $R_\textrm{P} = 1.37$ $R_\textrm{Jup}$ and cloud hole fraction $f_\textrm{hole} = 0.5$ are both larger than we found previously in \citep{Madurowicz_2023} which reported values of $R_\textrm{P} = 1.0$ $R_\textrm{Jup}$ and $f_\textrm{hole} = 0.2$. This can be attributed to the inclusion of the evolutionary prior in particular for the radius, which now enforces a radius $R_\textrm{p} \geq 1.35$ R$_\textrm{Jup}$. This also has the consequence of forcing the effective temperature to lower values to maintain the total luminosity. The larger cloud hole fraction could plausibly result from the interaction between opacity sources, as the newest best fit has a higher metallicity of around 0.5 dex, while metallicity was not at all considered in our previous analysis.
This best fit model is used to estimate the covariance on the errorbars for the NIRSpec data in the following manner. For reference, in previous examples applying this technique on IFU data \citep{Ruffio_2024}, we estimate the error covariance by computing the correlation profile from ``noise spectra," that is, spectra extracted from regions of the observation dataset with similar separations but different position angles than the extracted spectra from the region of the planet. However, with NIRSpec in the slit observing mode, the slit itself functionally obstructs these regions which would have similar noise properties, so we estimate the corelation profile from the model-subtracted data as a proxy for the noise spectra.
\begin{equation}
    \Psi_{\Delta l} = \langle r_l, r_{l + \Delta l} \rangle_l
\end{equation}
\noindent This is just a restatement of Equation (E1) in \citep{Ruffio_2024} with the average over spatial positions removed (since we only have one realization of the noise spectra instead of many). This results in a correlation profile which is decidedly more noisy. The correlation profile is shown in Figure (\ref{fig:correlation}).

An important observation from this figure is that the 1-2 pixel wide correlation which results from the regular wavelength grid interpolation step of the BREADS chain is still measurable, even with only a single instance of noise spectra. However, the typically seen 20-200 pixel scale oscillations which are interpreted as correlated errors from the spline subtraction are not visible. These errors are almost certainly present, but without more available noise spectra to estimate their magnitude, remain unmodeled and contribute to the errorbar inflation $\hat{e}$. However, the density of our spline nodes are relatively high, which contributes to mitigating their maximum extent.  In order to prevent this noisy correlation profile from affecting the covariance matrix, we enforce $\Psi(\Delta l \geq 3 \textrm{px}) = 0$ to ensure the stability of the matrix inversion. This curve can be seen with the dashed black line in the figure.
This covariance should be an improvement over ignoring the correlated errors completely, as it accounts for the part of the known error correlation due to the regular wavelength grid interpolation. This covariance model is used in the second iteration of model fitting, which also employs sub-model-grid interpolation to better sample the likelihood contours in the region around the best fit model.

\begin{figure}[H]
    \centering
    \includegraphics[width=1\linewidth]{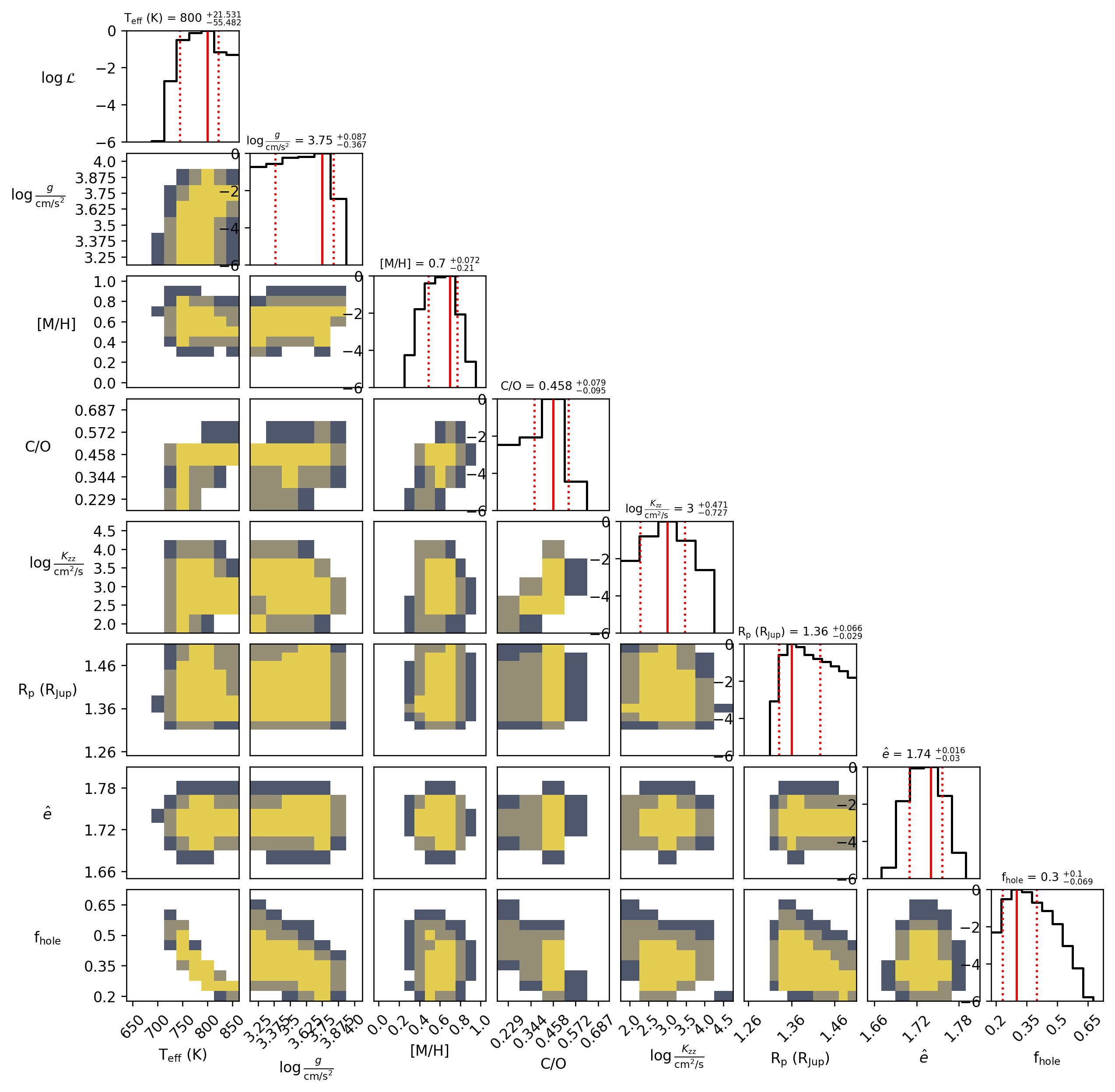}
    \caption{Triangle shaped posterior probability visualization for the second iteration of the atmospheric model likelihood calculation. A discrete colormap is used in the covariance plots with edges located at $\log \mathcal{L} \in [0, -2, -4, -6]$ so that the different colored regions in the diagram can be considered to be 1-, 2-, and 3-$\sigma$ regions. The mode and 1-$\sigma$ asymmetric limits of the single parameter marginal distributions are plotted with red lines and printed correspondingly in each title.}
    \label{fig:triangle}
\end{figure}

We perform the sub-grid interpolation on the discrete parameter vectors defined in the last column of Table (\ref{tab:parameter_iteration})
using the module \texttt{breads.atm\_utils.broadRGI} which internally relies on \texttt{scipy.RegularGridInterpolator} to perform the multi-linear interpolation over the input parameter space. While this methodology is rather commonplace it is also known to be problematic and could potentially be improved by replacing it with a more sophisticated interpolation scheme such as STARFISH \citep{Czekala_2015}. The posterior distributions calculated during the second iteration of the likelihood calculation are shown in Figure (\ref{fig:triangle}).

One observation to note is that the best fit model from the second iteration has slightly different parameter values when compared with the first iteration. In this case, we find that $T_\mathrm{eff}$ = 800 K, $\log g$ = 3.75, $[\mathrm{M}/\mathrm{H}]$ = 0.7, $\textrm{C}/\textrm{O}$ = 0.458, $\log K_\mathrm{zz}$ = 3, $R_\mathrm{P}$ = 1.36, $\hat{e}$ = 1.74, and $f_\mathrm{hole}$ = 0.3. Overall, the temperature, gravity, metallicity, C/O ratio, and eddy diffusion coefficient are all larger while the hole fraction is smaller. While the general structure of the atmosphere models are nearly indistinguishable by eye, this model predicts fainter values for the mid IR photometry, which may be more accurate given the limits set by \citet{Balmer_2025}. From the first iteration, our synthetic photometry of the model predicts a flux value that would have been detected at 2.87$\sigma$ in F430M, while the second iteration predicts instead a detection at 2.24$\sigma$. While both of these are below the 3$\sigma$ threshold that would be prudent to claim a detection, the first is closer to that threshold whereas the second is more marginal. Neither model predicts a significant detection in F460M, with an expectation at 1.05$\sigma$ and 0.79$\sigma$ for the first and second iterations respectively.

Regardless, both iterations have substantial preference for models with heightened metallicity relative to solar abundances. The first iteration prefers [M/H] = 0.5, while the second instead prefers [M/H] = 0.7, but with a 1$\sigma$ range that includes 0.5. This slightly higher metallicity is responsible for the darker color in F430M, as carbon dioxide abundance is very sensitive to metallicity, which is the dominant absorber in this band. \citet{Balmer_2025} did not report flux measurements for these bands, instead opting to report the more conservative upper limit. It is possible that absolute flux measurements at the planet's position in JWST/NIRCam F430M photometry could provide some additional discriminatory power over the reported upper limits, even if those flux measurements have large errorbars.

While the full triangle shaped posterior is useful to evaluate the possible covariances between model parameters, it is also interesting to ask how the posterior distributions for those model parameters depending on which datasets you choose to include in the calculation of the Likelihood. Equation (\ref{eq:chi2}) which computes the $\chi^2$ has three primary terms $\chi^2_\textrm{GPI}$, $\chi^2_\textrm{phot}$, and $\chi^2_\textrm{NIRSpec}$ which we sum to compute the total $\chi^2$ which then is proportional to the log likelihood. We could have instead chosen different subsets of these datasets to include in this computation, which effectively results in performing the same inference problem on different combinations of the data. Since there are three primary categories of data, there are seven possible combinations of these datasets we consider:
\begin{align}
    \log\mathcal{L}_\textrm{all} \propto - 1/2(\chi^2_\textrm{GPI} + \chi^2_\textrm{phot}+\chi^2_\textrm{NIRSpec}) + N_\mathrm{NIRSpec}/\hat{e} \\
    \log\mathcal{L}_\textrm{GPI} \propto - 1/2(\chi^2_\textrm{GPI} \color{white}+ \chi^2_\textrm{phot}+\chi^2_\textrm{NIRSpec}\color{black}) \color{white} + N_\mathrm{NIRSpec}/\hat{e}\color{black} \\
    \log\mathcal{L}_\textrm{phot} \propto - 1/2(\color{white}\chi^2_\textrm{GPI} +\color{black}\chi^2_\textrm{phot}\color{white}+\chi^2_\textrm{NIRSpec}\color{black})\color{white} + N_\mathrm{NIRSpec}/\hat{e}\color{black} \\
    \log\mathcal{L}_\textrm{NIRSpec} \propto - 1/2(\color{white}\chi^2_\textrm{GPI} + \chi^2_\textrm{phot}+\color{black}\chi^2_\textrm{NIRSpec}) + N_\mathrm{NIRSpec}/\hat{e} \\
    \log\mathcal{L}_\textrm{GPI+phot} \propto - 1/2(\chi^2_\textrm{GPI} + \chi^2_\textrm{phot}\color{white}+\chi^2_\textrm{NIRSpec}\color{black})\color{white}+ N_\mathrm{NIRSpec}/\hat{e}\color{black} \\
    \log\mathcal{L}_\textrm{GPI+NIRSpec} \propto - 1/2(\chi^2_\textrm{GPI} \color{white}+ \chi^2_\textrm{phot}
\color{black}+\chi^2_\textrm{NIRSpec}) +N_\mathrm{NIRSpec}/\hat{e} \\
    \log\mathcal{L}_\textrm{phot+NIRSpec} \propto - 1/2(\color{white}\chi^2_\textrm{GPI} + 
\color{black}\chi^2_\textrm{phot}+\chi^2_\textrm{NIRSpec}) +N_\mathrm{NIRSpec}/\hat{e} \\
\end{align}
\begin{figure}[H]
    \centering
    \includegraphics[width=1\linewidth]{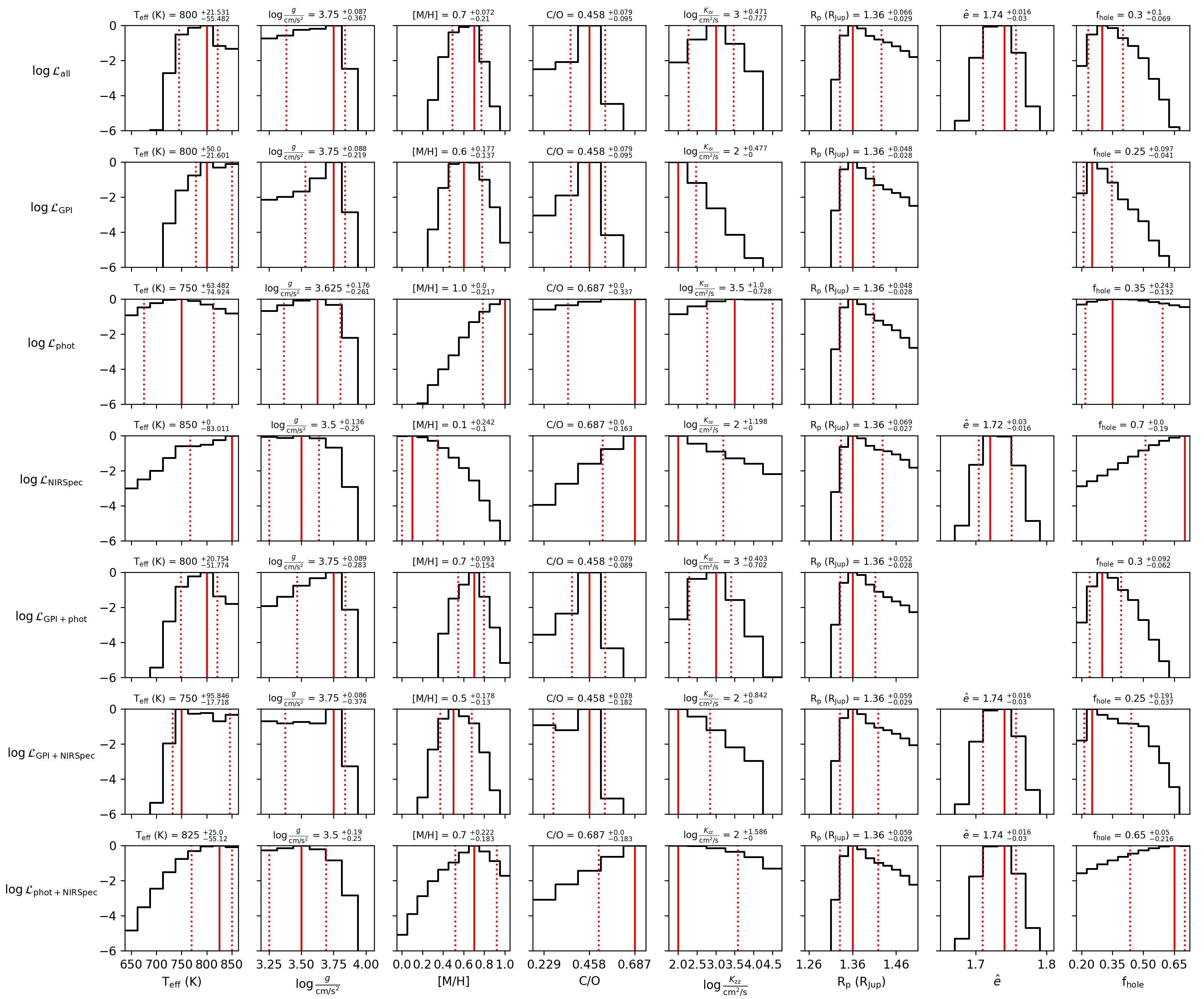}
    \caption{Single parameter marginalized posterior distributions for different combinations of datasets included the calculation of the likelihood. There are seven possible combinations of these datasets. From top to bottom: All datasets, just GPI spectroscopy, just photometry, just NIRSpec spectroscopy, GPI spectroscopy and all photometry, GPI spectroscopy and NIRSpec spectroscopy, and all photometry and NIRSpec spectroscopy. The $\hat{e}$ posterior is missing from the subsets which do not include NIRSpec.}
    \label{fig:posteriors}
\end{figure}
It is important to remark that the additional term $N_\textrm{NIRSpec}/\hat{e}$ only appears in combinations where the NIRSpec data is relevant, otherwise the $\hat{e}$ parameter is meaningless since it only applies to the NIRSpec data. For each of these combinations of datasets, we compute the single parameter marginal posterior distributions and show them for comparison in Figure (\ref{fig:posteriors}).

One of the observations of note within this figure is the tension between the photometry and the NIRSpec posteriors, in particular for temperature, metallicity and cloud hole fraction.  On its own, the NIRSpec data seems to prefer a higher effective temperature, lower metallicity, and larger cloud hole fraction, all of which together imply that the NIRSpec data on its own is better fit by a model which is brighter than would be expected when comparing with the photometry. Ultimately, in our likelihood framework, we assumed some unmodeled errorbar inflation $\hat{e}$ on the NIRSpec data, to deal with our inability to model the error covariance properly. However, this may have the consequence of also giving a degree of freedom to ignore what the NIRSpec data is telling the model would be best. One can speculate that if instead we had parameterized our likelihood with a floating offset parameter on the photometry as a plausible way to instead correct for inter-instrument systematics, we might see a substantially different answer for the best fit model. 

The cloud hole fraction degree of freedom gives the model substantial power to adjust the relative intensity of the planet brightness in the near IR (1-3 microns) relative to the mid IR (3-5 microns) because the cloud opacity dominates at the shorter wavelengths, which is easily seen in Figure (2) of \citet{Madurowicz_2023}. Since the prior we assumed for this analysis (Figure (\ref{fig:diamondback}) has a relatively narrow distribution for the planet radius, the NIRSpec only posterior is constrained into other parameters which could inflate the total brightness of the planet at these wavelengths, notably temperature, metallicity, and hole fraction.

\section{Discussion}

We have demonstrated the capabilities of an uncommon observing mode of JWST / NIRSpec. By using moderate resolution (R $\sim$ 2700) spectroscopy combined with software based starlight suppression techniques, we are able to detect a signal from 51 Eridani b at 4.8$\sigma$ in 3.43 hours of exposure time ($\sim$10 hours including overheads) without using a coronagraph. The planet is separated by $\sim$0.3 arcsec from its K = 4.5 magnitude host star with a global (planet-star) contrast ratio of C=1.1$\times10^{-5}$ and a local (planet-speckle) contrast ratio of Q=1.4$\times 10^{-3}$, the most challenging target we have detected with this technique. The signal is localized in the slit spatial coordinate (IFU Y) - planetary radial velocity phase space at the expected position of the planet. The peak of the of the cross-correlation function is spatially offset from the peak of the stellar flux, indicating that it is not an artifact of amplified stellar noise. The signal is absent (1.8$\sigma$) from the speckle-side pointing dataset. The signal is robust to analysis hyperparameters such as the node spacing, persisting at similar levels of confidence when reducing the data over a wide range of this parameter. ``Leave one molecule out" detection tests indicate this signal is primarily driven by the presence of methane ($\Delta\sigma_\textrm{CH$_4$} = -3.43$) and carbon monoxide ($\Delta\sigma_\textrm{CO} = -1.81$), which are expected from theoretical models to be the most prominent absorbers at these wavelengths and do not exist in the atmosphere of the stellar host. All of these lines of evidence and testing provide reassurance that the signal is indeed of planetary origin. Additionally, this detection is the first direct evidence of both molecules (CO + CH$_4$) coexisting in the planet's atmosphere, a signature of disequilibrium chemistry due to turbulent vertical mixing. The presence of disequilibrium chemistry has been suspected since the original discovery of methane \citep{Macintosh2015} in its atmosphere, and with our current detection predicated on the existence of CO in the planet's atmosphere, equilibrium chemistry is no longer a viable explanation. This is in contrast to the previous analysis in  \citet{Madurowicz_2023}, which did not have enough statistical evidence to rule out equilibrium chemistry models using only the single photometric point at M band.

However, our analysis is not perfect and our methods could be further refined and improved. Our comprehensive atmospheric model analysis indicates that the best fit NIRSpec errorbar inflation parameter $\hat{e} \sim 1.75$, which indicates that the photon noise errorbars on the data taken from the JWST pipeline underestimate the total error by roughly a factor of two. The primary source of the extra noise is almost certainly systematic error from the spline-based continuum subtraction which is known to result in spectrally correlated errors from previous work, although additional contributions to the error could come from imperfect calibrations of wavelength and spectral curvature in the NIRSpec instrument model used by the JWST pipeline. Future work on improving the spline modeling and continuum subtraction in general could reduce the systematic error term. In the context of doing similar observations with the NIRSpec IFU instead of the fixed slit, the covariance can be estimated from the data directly. By sampling the ``noise spectra" at spatial positions with similar flux levels as the position of the planet (the ``noise annulus") and computing their correlation profile, we can estimate the level of this covariance and include that model in our calculation of the likelihood. However, the primary disadvantage of using the fixed slit is that it does not sample those particular representative spatial positions, making this component of the covariance difficult to estimate. It is possible that with further work on this front, it could be possible to build a better covariance model from initial principles, with blind testing on different datasets with known estimations for validation. Application of such a covariance model could provide improvements in the computation of the cross correlation function and the full model likelihood, potentially resolving some discrepancies we observed or increasing the detection significance. We leave this question open for future inquiry. It is also possibly worth considering alternative modes of spline-subtraction which build into the model knowledge of the three dimensional structure of the JWST PSF, instead of our simple one dimensional treatment of the detector rows.

There is another downside to our observing strategy which warrants mention. Using the fixed slit instead of the IFU requires an additional pointing, the ``speckle-side" dataset, to compute the stellar spectrum and provide a null hypothesis test. This inflates the observing time required at a fixed SNR by a factor of two as the IFU observation strategy can effectively perform both pointings simultaneously. Our observations were planned with the fixed slit originally because of concerns with the detector saturating due to the extremely bright nature of the target star, however, with additional experience using these modes, we have begun to develop software strategies to mitigate detector saturation and charge bleeding from those saturated pixels. This program was designed during cycle 1, before the initial demonstration of JWST/NIRSpec in this mode on HD 19467 B had been finalized. While we have shown that using the fixed slit arrangement is a viable observing strategy, we would now recommend using the IFU over the fixed slit for the majority of high contrast spectroscopic targets. The charge bleeding effect can be mitigated as long as the planet is in the IFU slice adjacent to the primary slice which the star is contaminating. This was demonstrated in the Cycle 1 analysis on HD 19467 B \citep{Ruffio_2024} using a spline-based fit to the dark regions between the IFU traces. Improvements to this charge bleed modeling and subtraction using a Lorentzian-convolution subtraction are explored in (Ruffio \& Xuan 2025 in prep.) and continued to be investigated with further work on improving the analysis strategies.

Another interesting open question remains -- what would be the best strategy for optimal follow up observations of 51 Eridani b? One molecule that we searched for but only found a ``consistent non-detection" for is carbon dioxide (CO$_2$). The trade off between doing a deep F430M search with NIRCam to search for the depth of the CO$_2$ feature and further refine the constraint on the planet's metallicity could be compared with a longer spectroscopic integration using NIRSpec. How do both the absolute flux continuum level and total fractional bandpass coverage for a molecule compete in the effective detection SNR landscape as a function of exposure time? While NIRcam data was able to infer the existence of this molecule from a non-detection, our model fitting exercise indicates this measurement is possibly close to the edge of detectability, with expected model fluxes in the $2$-$3\sigma$ range in this band at the current NIRCam sensitivity. However, direct measurement of the CO absorption band with NIRCam was not close to the threshold, $\sim1\sigma$. At the very least, the NIRCam and NIRSpec data are complementary in that they are sensitive to different qualitative aspects of the planet's spectra. While our analysis presents some tension between the measurements between these two datasets, it is possible this tension arises from the unmodeled error covariance in our data.

Overall, we are just beginning to explore the capabilities of this novel application of direct spectroscopy to exoplanets. Being able to directly detect the presence of carbon monoxide from disequilibrium chemistry is one highlight, and the possibility to explore this technique at other wavelengths searching for other molecules and their implications is an unexplored frontier. While our software analysis techniques continue to be improved, it is possible that future reanalysis of these same datasets could provide improved results.

\section{Acknowledgments}
In no particular order, A.M. would like to thank William Balmer, Mathilde Malin, Sagnick Mukherjee, Natasha Batalha, Caroline Morley, and Laurent Pueyo for their helpful conversations. This work is based in part on observations made with the NASA/ESA/CSA James Webb Space Telescope. Support for program JWST-GO-03522 (A.M., J.-B.R.) was provided by NASA through grant from the Space Telescope Science Institute, which is operated by the Association of Universities for Research in Astronomy, Inc., under NASA contract NAS 5-03127. The data were obtained from the Mikulski Archive for Space Telescopes at the Space Telescope Science Institute. This work benefited from the 2024 Exoplanet Summer program in the Other Worlds Laboratory (OWL) at the University of California, Santa Cruz, a program funded by the Heising-Simons Foundation.

\section{Data/Code Availability}
The JWST/NIRSpec data analyzed in this paper is publicly available in MAST, as well as the raw JWST/NIRCam data. The auto-reduced NIRSpec datasets can be referenced though the DOI \url{http://dx.doi.org/10.17909/zj2p-hf65}, while our code contains a script to obtain the level 1 uncalibrated data products using the helper tool \texttt{jwst\_mast\_query.py}. The reduced GPI, JWST/NIRSpec, and JWST/NIRCam data products, as well as the scripts used to analyze the data and produce the figures in this paper, are available at the following repository:
\url{https://github.com/muuud/51-Eri-b-NIRSpec} or in Zenodo at the following DOI \url{http://dx.doi.org/10.5281/zenodo.17101634}.

\section{Software}
BREADS \citep{Agrawal_2023}, JWST \citep{jwst_pipeline}, PICASO \citep{Batalha2019}, VIRGA \citep{batalha2025}, astropy \citep{astropy:2013,astropy:2018,astropy:2022}, species \citep{Stolker2020}, matplotlib \citep{Hunter:2007}, numpy \citep{harris2020array}, scipy \citep{2020SciPy}

\section{Appendix: Injection and Recovery Test}

\begin{figure}[H]
    \centering
    \includegraphics[width=1\linewidth]{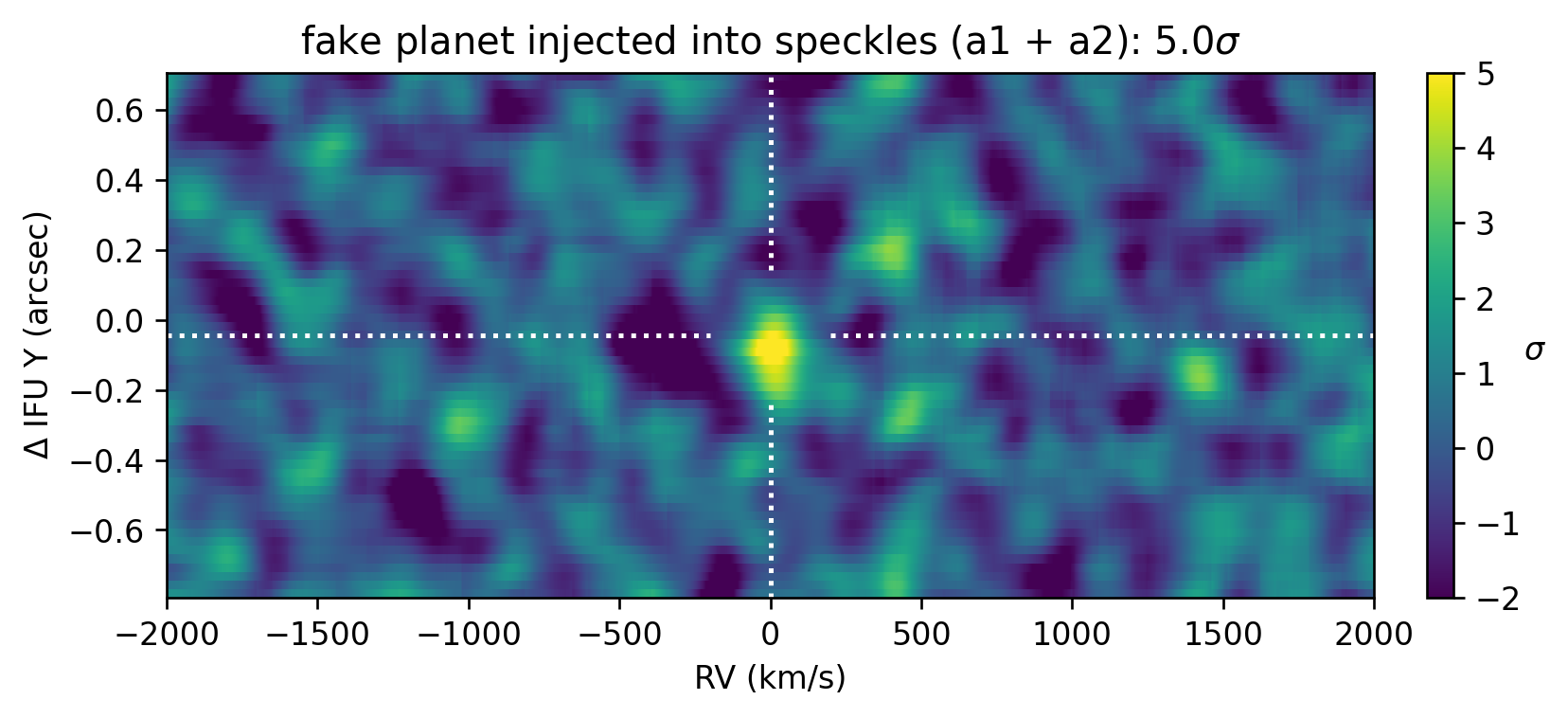}
    \caption{Artificial planetary signal injected into the continuum-subtracted point cloud for the speckle-side dataset is recovered in the CCF at a similar level of significance to the detection of 51 Eridani b in the planet-side dataset.}
    \label{fig:injection_recovery}
\end{figure}

In order to provide another line of evidence for the detection of 51 Eridani b, we perform an artificial injection and recovery test using only the speckle-side datasets. We inject a spline-filtered Elf-Owl model with parameters $T_\mathrm{eff} = 750$ K, $\log \big( \frac{g}{\mathrm{cm}/\mathrm{s}^2} \big) = $ 3.5, $[\mathrm{M}/\mathrm{H}] = 0.5$, $\textrm{C}/\textrm{O} = 0.458$, and $\log \big(\frac{K_\mathrm{zz}}{\mathrm{cm}^2\mathrm{s}}\big) = 2$, into the speckle-side continuum-subtracted point clouds using a synthetic WebbPSF interpolation after converting the model into the detector units of MJy. Then we perform the spectral extraction and cross-correlation analysis on the perturbed speckle-side point clouds following the same procedure as the main analysis. The resulting phase space CCF from this injection and recovery test is shown in Figure (\ref{fig:injection_recovery}) demonstrating that the fake signal is recovered at a significance of $5.0\sigma$

This detection significance is very similar to but slightly higher than the real dataset, which could be plausibly attributed to data-model mismatch, as the injected signal is a perfect match to the CCF template used in this analysis which is not necessarily true of the real data. Additionally, the speckle-side only residual noise does not contain minor errors from spectral mismatch that the planet-side dataset has. This spectral mismatch could arise from the difference in the slit illumination profile between the planet- and speckle-side datasets.

\section{Appendix: Additional Figures}
\begin{figure}[H]
    \centering
    \includegraphics[width=1\linewidth]{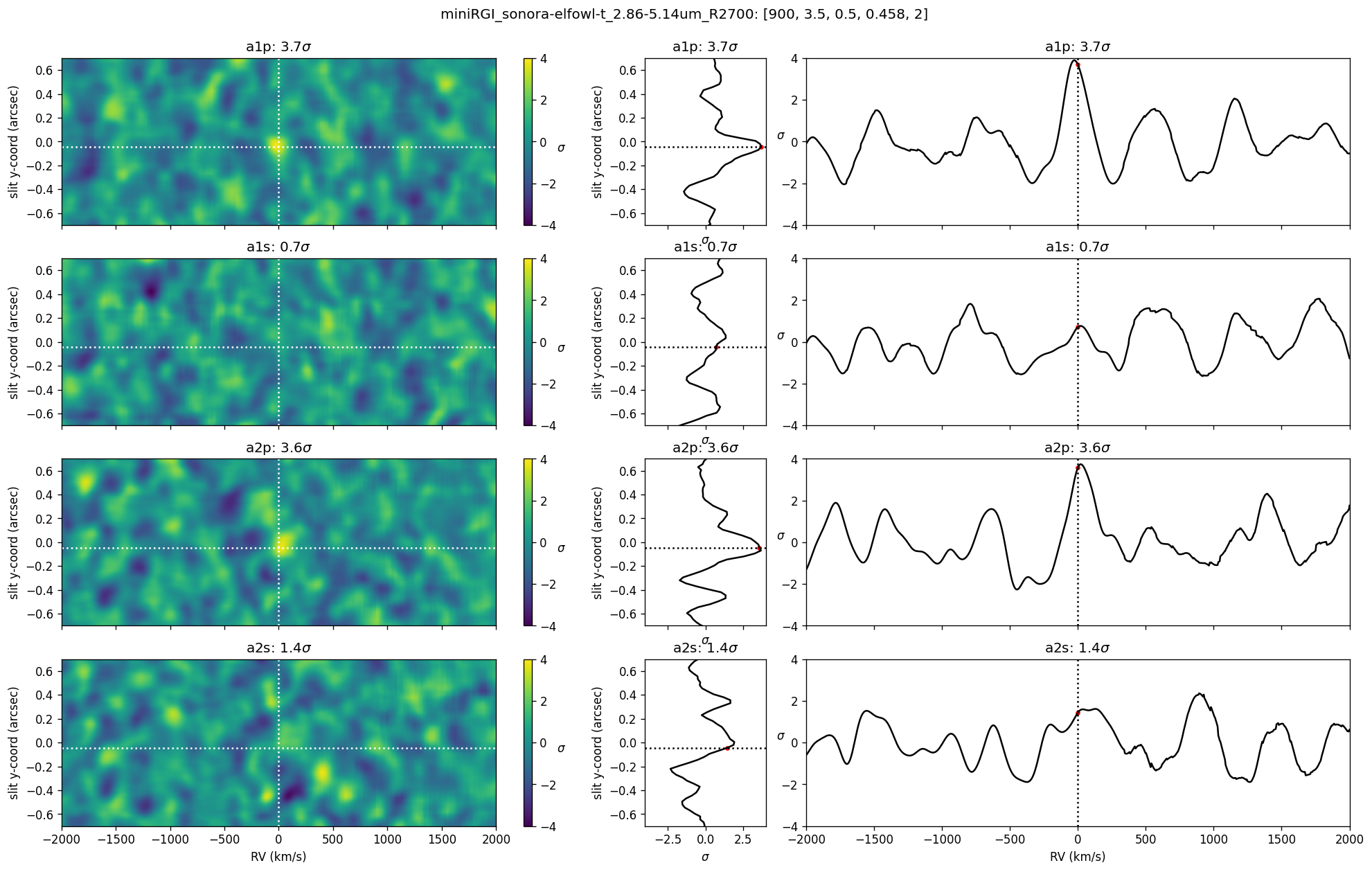}
    \caption{Cross correlation function calculated on individual datasets, showing  detections in slits A1 and A2 individually. (Left) 2D CCF detection maps in IFU Y - RV phase space, (Middle) CCF slice through RV = 0 km/s, (Right) CCF slice through IFU Y = -0.04 arcsec, the expected position of the planet.}
    \label{fig:fullCCF}
\end{figure}

\bibliography{main}{}

\begin{thebibliography}{}
\expandafter\ifx\csname natexlab\endcsname\relax\def\natexlab#1{#1}\fi
\providecommand{\url}[1]{\href{#1}{#1}}
\providecommand{\dodoi}[1]{doi:~\href{http://doi.org/#1}{\nolinkurl{#1}}}
\providecommand{\doeprint}[1]{\href{http://ascl.net/#1}{\nolinkurl{http://ascl.net/#1}}}
\providecommand{\doarXiv}[1]{\href{https://arxiv.org/abs/#1}{\nolinkurl{https://arxiv.org/abs/#1}}}

\bibitem[{Agrawal {et~al.}(2023)Agrawal, Ruffio, Konopacky, Macintosh, Mawet, Nielsen, Hoch, Liu, Barman, Thompson, Greenbaum, Marois, \& Patience}]{Agrawal_2023}
Agrawal, S., Ruffio, J.-B., Konopacky, Q.~M., {et~al.} 2023, The Astronomical Journal, 166, 15, \dodoi{10.3847/1538-3881/acd6a3}

\bibitem[{Alam {et~al.}(2024)Alam, Gao, Adams~Redai, Wallack, Wogan, Aguichine, Dattilo, Alderson, Batalha, Batalha, Kirk, López-Morales, Meech, Moran, Teske, Wakeford, \& Wolfgang}]{Alam_2025}
Alam, M.~K., Gao, P., Adams~Redai, J., {et~al.} 2024, The Astronomical Journal, 169, 15, \dodoi{10.3847/1538-3881/ad8eb5}

\bibitem[{Alderson {et~al.}(2024)Alderson, Batalha, Wakeford, Wallack, Aguichine, Teske, Adams~Redai, Alam, Batalha, Gao, Kirk, López-Morales, Moran, Scarsdale, Wogan, \& Wolfgang}]{Alderson_2024}
Alderson, L., Batalha, N.~E., Wakeford, H.~R., {et~al.} 2024, The Astronomical Journal, 167, 216, \dodoi{10.3847/1538-3881/ad32c9}

\bibitem[{Alderson {et~al.}(2025)Alderson, Moran, Wallack, Batalha, Wogan, Dattilo, Wakeford, Redai, Alam, Aguichine, Batalha, Gagnebin, Gao, Kirk, López-Morales, Meech, Teske, \& Wolfgang}]{Alderson_2025}
Alderson, L., Moran, S.~E., Wallack, N.~L., {et~al.} 2025, The Astronomical Journal, 169, 142, \dodoi{10.3847/1538-3881/adad64}

\bibitem[{{Astropy Collaboration} {et~al.}(2013){Astropy Collaboration}, {Robitaille}, {Tollerud}, {Greenfield}, {Droettboom}, {Bray}, {Aldcroft}, {Davis}, {Ginsburg}, {Price-Whelan}, {Kerzendorf}, {Conley}, {Crighton}, {Barbary}, {Muna}, {Ferguson}, {Grollier}, {Parikh}, {Nair}, {Unther}, {Deil}, {Woillez}, {Conseil}, {Kramer}, {Turner}, {Singer}, {Fox}, {Weaver}, {Zabalza}, {Edwards}, {Azalee Bostroem}, {Burke}, {Casey}, {Crawford}, {Dencheva}, {Ely}, {Jenness}, {Labrie}, {Lim}, {Pierfederici}, {Pontzen}, {Ptak}, {Refsdal}, {Servillat}, \& {Streicher}}]{astropy:2013}
{Astropy Collaboration}, {Robitaille}, T.~P., {Tollerud}, E.~J., {et~al.} 2013, \aap, 558, A33, \dodoi{10.1051/0004-6361/201322068}

\bibitem[{{Astropy Collaboration} {et~al.}(2018){Astropy Collaboration}, {Price-Whelan}, {Sip{\H{o}}cz}, {G{\"u}nther}, {Lim}, {Crawford}, {Conseil}, {Shupe}, {Craig}, {Dencheva}, {Ginsburg}, {Vand erPlas}, {Bradley}, {P{\'e}rez-Su{\'a}rez}, {de Val-Borro}, {Aldcroft}, {Cruz}, {Robitaille}, {Tollerud}, {Ardelean}, {Babej}, {Bach}, {Bachetti}, {Bakanov}, {Bamford}, {Barentsen}, {Barmby}, {Baumbach}, {Berry}, {Biscani}, {Boquien}, {Bostroem}, {Bouma}, {Brammer}, {Bray}, {Breytenbach}, {Buddelmeijer}, {Burke}, {Calderone}, {Cano Rodr{\'\i}guez}, {Cara}, {Cardoso}, {Cheedella}, {Copin}, {Corrales}, {Crichton}, {D'Avella}, {Deil}, {Depagne}, {Dietrich}, {Donath}, {Droettboom}, {Earl}, {Erben}, {Fabbro}, {Ferreira}, {Finethy}, {Fox}, {Garrison}, {Gibbons}, {Goldstein}, {Gommers}, {Greco}, {Greenfield}, {Groener}, {Grollier}, {Hagen}, {Hirst}, {Homeier}, {Horton}, {Hosseinzadeh}, {Hu}, {Hunkeler}, {Ivezi{\'c}}, {Jain}, {Jenness}, {Kanarek}, {Kendrew}, {Kern}, {Kerzendorf}, {Khvalko}, {King}, {Kirkby}, {Kulkarni},
  {Kumar}, {Lee}, {Lenz}, {Littlefair}, {Ma}, {Macleod}, {Mastropietro}, {McCully}, {Montagnac}, {Morris}, {Mueller}, {Mumford}, {Muna}, {Murphy}, {Nelson}, {Nguyen}, {Ninan}, {N{\"o}the}, {Ogaz}, {Oh}, {Parejko}, {Parley}, {Pascual}, {Patil}, {Patil}, {Plunkett}, {Prochaska}, {Rastogi}, {Reddy Janga}, {Sabater}, {Sakurikar}, {Seifert}, {Sherbert}, {Sherwood-Taylor}, {Shih}, {Sick}, {Silbiger}, {Singanamalla}, {Singer}, {Sladen}, {Sooley}, {Sornarajah}, {Streicher}, {Teuben}, {Thomas}, {Tremblay}, {Turner}, {Terr{\'o}n}, {van Kerkwijk}, {de la Vega}, {Watkins}, {Weaver}, {Whitmore}, {Woillez}, {Zabalza}, \& {Astropy Contributors}}]{astropy:2018}
{Astropy Collaboration}, {Price-Whelan}, A.~M., {Sip{\H{o}}cz}, B.~M., {et~al.} 2018, \aj, 156, 123, \dodoi{10.3847/1538-3881/aabc4f}

\bibitem[{{Astropy Collaboration} {et~al.}(2022){Astropy Collaboration}, {Price-Whelan}, {Lim}, {Earl}, {Starkman}, {Bradley}, {Shupe}, {Patil}, {Corrales}, {Brasseur}, {N{"o}the}, {Donath}, {Tollerud}, {Morris}, {Ginsburg}, {Vaher}, {Weaver}, {Tocknell}, {Jamieson}, {van Kerkwijk}, {Robitaille}, {Merry}, {Bachetti}, {G{"u}nther}, {Aldcroft}, {Alvarado-Montes}, {Archibald}, {B{'o}di}, {Bapat}, {Barentsen}, {Baz{'a}n}, {Biswas}, {Boquien}, {Burke}, {Cara}, {Cara}, {Conroy}, {Conseil}, {Craig}, {Cross}, {Cruz}, {D'Eugenio}, {Dencheva}, {Devillepoix}, {Dietrich}, {Eigenbrot}, {Erben}, {Ferreira}, {Foreman-Mackey}, {Fox}, {Freij}, {Garg}, {Geda}, {Glattly}, {Gondhalekar}, {Gordon}, {Grant}, {Greenfield}, {Groener}, {Guest}, {Gurovich}, {Handberg}, {Hart}, {Hatfield-Dodds}, {Homeier}, {Hosseinzadeh}, {Jenness}, {Jones}, {Joseph}, {Kalmbach}, {Karamehmetoglu}, {Ka{l}uszy{'n}ski}, {Kelley}, {Kern}, {Kerzendorf}, {Koch}, {Kulumani}, {Lee}, {Ly}, {Ma}, {MacBride}, {Maljaars}, {Muna}, {Murphy}, {Norman}, {O'Steen},
  {Oman}, {Pacifici}, {Pascual}, {Pascual-Granado}, {Patil}, {Perren}, {Pickering}, {Rastogi}, {Roulston}, {Ryan}, {Rykoff}, {Sabater}, {Sakurikar}, {Salgado}, {Sanghi}, {Saunders}, {Savchenko}, {Schwardt}, {Seifert-Eckert}, {Shih}, {Jain}, {Shukla}, {Sick}, {Simpson}, {Singanamalla}, {Singer}, {Singhal}, {Sinha}, {Sip{H{o}}cz}, {Spitler}, {Stansby}, {Streicher}, {{{S}}umak}, {Swinbank}, {Taranu}, {Tewary}, {Tremblay}, {Val-Borro}, {Van Kooten}, {Vasovi{'c}}, {Verma}, {de Miranda Cardoso}, {Williams}, {Wilson}, {Winkel}, {Wood-Vasey}, {Xue}, {Yoachim}, {Zhang}, {Zonca}, \& {Astropy Project Contributors}}]{astropy:2022}
{Astropy Collaboration}, {Price-Whelan}, A.~M., {Lim}, P.~L., {et~al.} 2022, \apj, 935, 167, \dodoi{10.3847/1538-4357/ac7c74}

\bibitem[{Balmer {et~al.}(2025)Balmer, Kammerer, Pueyo, Perrin, Girard, Leisenring, Lawson, Dennen, van~der Marel, Beichman, Bryden, Llop-Sayson, Valenti, Lothringer, Lewis, Mâlin, Rebollido, Rickman, Hoch, Soummer, Clampin, \& Mountain}]{Balmer_2025}
Balmer, W.~O., Kammerer, J., Pueyo, L., {et~al.} 2025, The Astronomical Journal, 169, 209, \dodoi{10.3847/1538-3881/adb1c6}

\bibitem[{{Batalha} {et~al.}(2019){Batalha}, {Marley}, {Lewis}, \& {Fortney}}]{Batalha2019}
{Batalha}, N.~E., {Marley}, M.~S., {Lewis}, N.~K., \& {Fortney}, J.~J. 2019, \apj, 878, 70, \dodoi{10.3847/1538-4357/ab1b51}

\bibitem[{Batalha {et~al.}(2025)Batalha, Rooney, Visscher, Moran, Marley, Sengupta, Kiefer, Lodge, Mang, Morley, Mukherjee, Fortney, Gao, Lewis, Mayorga, Pearce, \& Wakeford}]{batalha2025}
Batalha, N.~E., Rooney, C.~M., Visscher, C., {et~al.} 2025, Condensation Clouds in Substellar Atmospheres with Virga.
\newblock \doarXiv{2508.15102}

\bibitem[{Benneke {et~al.}(2024)Benneke, Roy, Coulombe, Radica, Piaulet, Ahrer, Pierrehumbert, Krissansen-Totton, Schlichting, Hu, Yang, Christie, Thorngren, Young, Pelletier, Knutson, Miguel, Evans-Soma, Dorn, Gagnebin, Fortney, Komacek, MacDonald, Raul, Cloutier, Acuna, Lafrenière, Cadieux, Doyon, Welbanks, \& Allart}]{benneke2024}
Benneke, B., Roy, P.-A., Coulombe, L.-P., {et~al.} 2024, JWST Reveals CH$_4$, CO$_2$, and H$_2$O in a Metal-rich Miscible Atmosphere on a Two-Earth-Radius Exoplanet.
\newblock \doarXiv{2403.03325}

\bibitem[{{Birkmann, S. M.} {et~al.}(2022){Birkmann, S. M.}, {Ferruit, P.}, {Giardino, G.}, {Nielsen, L. D.}, {García Muñoz, A.}, {Kendrew, S.}, {Rauscher, B. J.}, {Beck, T. L.}, {Keyes, C.}, {Valenti, J. A.}, {Jakobsen, P.}, {Dorner, B.}, {Alves de Oliveira, C.}, {Arribas, S.}, {Böker, T.}, {Bunker, A. J.}, {Charlot, S.}, {de Marchi, G.}, {Kumari, N.}, {López-Caniego, M.}, {Lützgendorf, N.}, {Maiolino, R.}, {Manjavacas, E.}, {Marston, A.}, {Moseley, S. H.}, {Prizkal, N.}, {Proffitt, C.}, {Rawle, T.}, {Rix, H.-W.}, {te Plate, M.}, {Sabbi, E.}, {Sirianni, M.}, {Willott, C. J.}, \& {Zeidler, P.}}]{Birkmann2022}
{Birkmann, S. M.}, {Ferruit, P.}, {Giardino, G.}, {et~al.} 2022, A\&A, 661, A83, \dodoi{10.1051/0004-6361/202142592}

\bibitem[{Bogat {et~al.}(2025)Bogat, Schlieder, Lawson, Li, Leisenring, Meyer, Balmer, Barclay, Beichman, Bryden, Calissendorff, Carter, Furio, Girard, Greene, Groff, Kammerer, Llop-Sayson, McElwain, Rieke, \& Ygouf}]{bogat2025}
Bogat, E., Schlieder, J.~E., Lawson, K.~D., {et~al.} 2025, Probing the Outskirts of M Dwarf Planetary Systems with a Cycle 1 JWST NIRCam Coronagraphy Survey.
\newblock \doarXiv{2504.11659}

\bibitem[{{Brown-Sevilla, S. B.} {et~al.}(2023){Brown-Sevilla, S. B.}, {Maire, A.-L.}, {Mollière, P.}, {Samland, M.}, {Feldt, M.}, {Brandner, W.}, {Henning, Th.}, {Gratton, R.}, {Janson, M.}, {Stolker, T.}, {Hagelberg, J.}, {Zurlo, A.}, {Cantalloube, F.}, {Boccaletti, A.}, {Bonnefoy, M.}, {Chauvin, G.}, {Desidera, S.}, {D'Orazi, V.}, {Lagrange, A.-M.}, {Langlois, M.}, {Menard, F.}, {Mesa, D.}, {Meyer, M.}, {Pavlov, A.}, {Petit, C.}, {Rochat, S.}, {Rouan, D.}, {Schmidt, T.}, {Vigan, A.}, \& {Weber, L.}}]{BrownSevilla2023}
{Brown-Sevilla, S. B.}, {Maire, A.-L.}, {Mollière, P.}, {et~al.} 2023, A\&A, 673, A98, \dodoi{10.1051/0004-6361/202244826}

\bibitem[{Bushouse {et~al.}(2025)Bushouse, Eisenhamer, Dencheva, Davies, Greenfield, Morrison, Hodge, Simon, Grumm, Droettboom, Slavich, Sosey, Pauly, Miller, Jedrzejewski, Hack, Davis, Crawford, Law, Gordon, Regan, Cara, MacDonald, Bradley, Shanahan, Jamieson, Teodoro, Williams, Pena-Guerrero, Graham, Molter, Brandt, Hayes, Cooper, Clarke, \& Filippazzo}]{jwst_pipeline}
Bushouse, H., Eisenhamer, J., Dencheva, N., {et~al.} 2025, \dodoi{10.5281/zenodo.14597407}

\bibitem[{{Böker, T.} {et~al.}(2022){Böker, T.}, {Arribas, S.}, {Lützgendorf, N.}, {Alves de Oliveira, C.}, {Beck, T. L.}, {Birkmann, S.}, {Bunker, A. J.}, {Charlot, S.}, {de Marchi, G.}, {Ferruit, P.}, {Giardino, G.}, {Jakobsen, P.}, {Kumari, N.}, {López-Caniego, M.}, {Maiolino, R.}, {Manjavacas, E.}, {Marston, A.}, {Moseley, S. H.}, {Muzerolle, J.}, {Ogle, P.}, {Pirzkal, N.}, {Rauscher, B.}, {Rawle, T.}, {Rix, H.-W.}, {Sabbi, E.}, {Sargent, B.}, {Sirianni, M.}, {te Plate, M.}, {Valenti, J.}, {Willott, C. J.}, \& {Zeidler, P.}}]{Boker2022}
{Böker, T.}, {Arribas, S.}, {Lützgendorf, N.}, {et~al.} 2022, A\&A, 661, A82, \dodoi{10.1051/0004-6361/202142589}

\bibitem[{Carter {et~al.}(2020)Carter, Hinkley, Bonavita, Phillips, Girard, Perrin, Pueyo, Vigan, Gagné, \& Skemer}]{Carter2020}
Carter, A.~L., Hinkley, S., Bonavita, M., {et~al.} 2020, Monthly Notices of the Royal Astronomical Society, 501, 1999, \dodoi{10.1093/mnras/staa3579}

\bibitem[{Czekala {et~al.}(2015)Czekala, Andrews, Mandel, Hogg, \& Green}]{Czekala_2015}
Czekala, I., Andrews, S.~M., Mandel, K.~S., Hogg, D.~W., \& Green, G.~M. 2015, The Astrophysical Journal, 812, 128, \dodoi{10.1088/0004-637X/812/2/128}

\bibitem[{Delorme {et~al.}(2021)Delorme, Jovanovic, Echeverri, Mawet, Wallace, Bartos, Cetre, Wizinowich, Ragland, Lilley, Wetherell, Doppmann, Wang, Morris, Ruffio, Martin, Fitzgerald, Ruane, Schofield, Suominen, Calvin, Wang, Magnone, Johnson, Sohn, Lopez, Bond, Pezzato, Llop-Sayson, Chun, \& Skemer}]{Delorme2021}
Delorme, J.-R., Jovanovic, N., Echeverri, D., {et~al.} 2021, Journal of Astronomical Telescopes, Instruments, and Systems, 7, 035006, \dodoi{10.1117/1.JATIS.7.3.035006}

\bibitem[{Elliott {et~al.}(2024)Elliott, Boyajian, Ellis, von Braun, Mann, \& Schaefer}]{Elliott_2024}
Elliott, A., Boyajian, T., Ellis, T., {et~al.} 2024, Publications of the Astronomical Society of Australia, 41, e043, \dodoi{10.1017/pasa.2024.40}

\bibitem[{Espinoza {et~al.}(2023)Espinoza, Úbeda, Birkmann, Ferruit, Valenti, Sing, Rustamkulov, Regan, Kendrew, Sabbi, Schlawin, Beatty, Albert, Greene, Nikolov, Karakla, Keyes, Alves~de Oliveira, Böker, Pena-Guerrero, Giardino, Kumari, Manjavacas, Proffitt, \& Rawle}]{Espinoza_2023}
Espinoza, N., Úbeda, L., Birkmann, S.~M., {et~al.} 2023, Publications of the Astronomical Society of the Pacific, 135, 018002, \dodoi{10.1088/1538-3873/aca3d3}

\bibitem[{Gressier {et~al.}(2024)Gressier, Espinoza, Allen, Sing, Banerjee, Barstow, Valenti, Lewis, Birkmann, Challener, Manjavacas, Alves~de Oliveira, Crouzet, \& Beck}]{Gressier_2024}
Gressier, A., Espinoza, N., Allen, N.~H., {et~al.} 2024, The Astrophysical Journal Letters, 975, L10, \dodoi{10.3847/2041-8213/ad73d1}

\bibitem[{Harris {et~al.}(2020)Harris, Millman, van~der Walt, Gommers, Virtanen, Cournapeau, Wieser, Taylor, Berg, Smith, Kern, Picus, Hoyer, van Kerkwijk, Brett, Haldane, del R{\'{i}}o, Wiebe, Peterson, G{\'{e}}rard-Marchant, Sheppard, Reddy, Weckesser, Abbasi, Gohlke, \& Oliphant}]{harris2020array}
Harris, C.~R., Millman, K.~J., van~der Walt, S.~J., {et~al.} 2020, Nature, 585, 357, \dodoi{10.1038/s41586-020-2649-2}

\bibitem[{Hoch {et~al.}(2023)Hoch, Konopacky, Theissen, Ruffio, Barman, Rickman, Perrin, Macintosh, \& Marois}]{Hoch_2023}
Hoch, K. K.~W., Konopacky, Q.~M., Theissen, C.~A., {et~al.} 2023, The Astronomical Journal, 166, 85, \dodoi{10.3847/1538-3881/ace442}

\bibitem[{Hoch {et~al.}(2024)Hoch, Theissen, Barman, Perrin, Ruffio, Rickman, Konopacky, Manjavacas, Balmer, Pueyo, Kammerer, van~der Marel, Lewis, Girard, Seager, Clampin, \& Mountain}]{Hoch_2024}
Hoch, K. K.~W., Theissen, C.~A., Barman, T.~S., {et~al.} 2024, The Astronomical Journal, 168, 187, \dodoi{10.3847/1538-3881/ad6cd3}

\bibitem[{Hunter(2007)}]{Hunter:2007}
Hunter, J.~D. 2007, Computing in Science \& Engineering, 9, 90, \dodoi{10.1109/MCSE.2007.55}

\bibitem[{{Jakobsen, P.} {et~al.}(2022){Jakobsen, P.}, {Ferruit, P.}, {Alves de Oliveira, C.}, {Arribas, S.}, {Bagnasco, G.}, {Barho, R.}, {Beck, T. L.}, {Birkmann, S.}, {Böker, T.}, {Bunker, A. J.}, {Charlot, S.}, {de Jong, P.}, {de Marchi, G.}, {Ehrenwinkler, R.}, {Falcolini, M.}, {Fels, R.}, {Franx, M.}, {Franz, D.}, {Funke, M.}, {Giardino, G.}, {Gnata, X.}, {Holota, W.}, {Honnen, K.}, {Jensen, P. L.}, {Jentsch, M.}, {Johnson, T.}, {Jollet, D.}, {Karl, H.}, {Kling, G.}, {Köhler, J.}, {Kolm, M.-G.}, {Kumari, N.}, {Lander, M. E.}, {Lemke, R.}, {López-Caniego, M.}, {Lützgendorf, N.}, {Maiolino, R.}, {Manjavacas, E.}, {Marston, A.}, {Maschmann, M.}, {Maurer, R.}, {Messerschmidt, B.}, {Moseley, S. H.}, {Mosner, P.}, {Mott, D. B.}, {Muzerolle, J.}, {Pirzkal, N.}, {Pittet, J.-F.}, {Plitzke, A.}, {Posselt, W.}, {Rapp, B.}, {Rauscher, B. J.}, {Rawle, T.}, {Rix, H.-W.}, {Rödel, A.}, {Rumler, P.}, {Sabbi, E.}, {Salvignol, J.-C.}, {Schmid, T.}, {Sirianni, M.}, {Smith, C.}, {Strada, P.}, {te Plate, M.}, {Valenti,
  J.}, {Wettemann, T.}, {Wiehe, T.}, {Wiesmayer, M.}, {Willott, C. J.}, {Wright, R.}, {Zeidler, P.}, \& {Zincke, C.}}]{Jakobsen2022}
{Jakobsen, P.}, {Ferruit, P.}, {Alves de Oliveira, C.}, {et~al.} 2022, A\&A, 661, A80, \dodoi{10.1051/0004-6361/202142663}

\bibitem[{Luque {et~al.}(2024)Luque, {Park Coy}, Xue, Feinstein, Ahrer, Changeat, Zhang, Moran, Bean, Kite, {Weiner Mansfield}, \& Pall{\'e}}]{Luque2024}
Luque, R., {Park Coy}, B., Xue, Q., {et~al.} 2024, The Astronomical Journal, \dodoi{10.48550/arXiv.2412.03411}

\bibitem[{Macintosh {et~al.}(2015)Macintosh, Graham, Barman, Rosa, Konopacky, Marley, Marois, Nielsen, Pueyo, Rajan, Rameau, Saumon, Wang, Patience, Ammons, Arriaga, Artigau, Beckwith, Brewster, Bruzzone, Bulger, Burningham, Burrows, Chen, Chiang, Chilcote, Dawson, Dong, Doyon, Draper, Duchêne, Esposito, Fabrycky, Fitzgerald, Follette, Fortney, Gerard, Goodsell, Greenbaum, Hibon, Hinkley, Cotten, Hung, Ingraham, Johnson-Groh, Kalas, Lafreniere, Larkin, Lee, Line, Long, Maire, Marchis, Matthews, Max, Metchev, Millar-Blanchaer, Mittal, Morley, Morzinski, Murray-Clay, Oppenheimer, Palmer, Patel, Perrin, Poyneer, Rafikov, Rantakyrö, Rice, Rojo, Rudy, Ruffio, Ruiz, Sadakuni, Saddlemyer, Salama, Savransky, Schneider, Sivaramakrishnan, Song, Soummer, Thomas, Vasisht, Wallace, Ward-Duong, Wiktorowicz, Wolff, \& Zuckerman}]{Macintosh2015}
Macintosh, B., Graham, J.~R., Barman, T., {et~al.} 2015, Science, 350, 64, \dodoi{10.1126/science.aac5891}

\bibitem[{Madurowicz {et~al.}(2023)Madurowicz, Mukherjee, Batalha, Macintosh, Marley, \& Karalidi}]{Madurowicz_2023}
Madurowicz, A., Mukherjee, S., Batalha, N., {et~al.} 2023, The Astronomical Journal, 165, 238, \dodoi{10.3847/1538-3881/acca7a}

\bibitem[{Marley {et~al.}(2010)Marley, Saumon, \& Goldblatt}]{Marley2010}
Marley, M.~S., Saumon, D., \& Goldblatt, C. 2010, The Astrophysical Journal Letters, 723, L117, \dodoi{10.1088/2041-8205/723/1/L117}

\bibitem[{Morley {et~al.}(2024{\natexlab{a}})Morley, Mukherjee, Marley, Fortney, Visscher, Lupu, Gharib-Nezhad, Thorngren, Freedman, \& Batalha}]{morley_2024_zenodo}
Morley, C., Mukherjee, S., Marley, M., {et~al.} 2024{\natexlab{a}}, The Sonora Substellar Atmosphere Models: Diamondback,  Zenodo, \dodoi{10.5281/zenodo.12735103}

\bibitem[{Morley {et~al.}(2024{\natexlab{b}})Morley, Mukherjee, Marley, Fortney, Visscher, Lupu, Gharib-Nezhad, Thorngren, Freedman, \& Batalha}]{Morley_2024}
Morley, C.~V., Mukherjee, S., Marley, M.~S., {et~al.} 2024{\natexlab{b}}, The Astrophysical Journal, 975, 59, \dodoi{10.3847/1538-4357/ad71d5}

\bibitem[{Mukherjee {et~al.}(2023)Mukherjee, Fortney, Morley, Batalha, Marley, Karalidi, Visscher, Lupu, Freedman, \& Gharib-Nezhad}]{mukherjee_2023}
Mukherjee, S., Fortney, J., Morley, C., {et~al.} 2023, The Sonora Substellar Atmosphere Models. IV. Elf Owl: Atmospheric Mixing and Chemical Disequilibrium with Varying Metallicity and C/O Ratios (T- type Models),  Zenodo, \dodoi{10.5281/zenodo.10385821}

\bibitem[{Mukherjee {et~al.}(2024)Mukherjee, Fortney, Morley, Batalha, Marley, Karalidi, Visscher, Lupu, Freedman, \& Gharib-Nezhad}]{Mukherjee_2024}
Mukherjee, S., Fortney, J.~J., Morley, C.~V., {et~al.} 2024, The Astrophysical Journal, 963, 73, \dodoi{10.3847/1538-4357/ad18c2}

\bibitem[{Nielsen {et~al.}(2019)Nielsen, De~Rosa, Macintosh, Wang, Ruffio, Chiang, Marley, Saumon, Savransky, Mark~Ammons, Bailey, Barman, Blain, Bulger, Burrows, Chilcote, Cotten, Czekala, Doyon, Duchêne, Esposito, Fabrycky, Fitzgerald, Follette, Fortney, Gerard, Goodsell, Graham, Greenbaum, Hibon, Hinkley, Hirsch, Hom, Hung, Ilene~Dawson, Ingraham, Kalas, Konopacky, Larkin, Lee, Lin, Maire, Marchis, Marois, Metchev, Millar-Blanchaer, Morzinski, Oppenheimer, Palmer, Patience, Perrin, Poyneer, Pueyo, Rafikov, Rajan, Rameau, Rantakyrö, Ren, Schneider, Sivaramakrishnan, Song, Soummer, Tallis, Thomas, Ward-Duong, \& Wolff}]{Nielsen_2019}
Nielsen, E.~L., De~Rosa, R.~J., Macintosh, B., {et~al.} 2019, The Astronomical Journal, 158, 13, \dodoi{10.3847/1538-3881/ab16e9}

\bibitem[{Rajan {et~al.}(2017)Rajan, Rameau, Rosa, Marley, Graham, Macintosh, Marois, Morley, Patience, Pueyo, Saumon, Ward-Duong, Ammons, Arriaga, Bailey, Barman, Bulger, Burrows, Chilcote, Cotten, Czekala, Doyon, Duchêne, Esposito, Fitzgerald, Follette, Fortney, Goodsell, Greenbaum, Hibon, Hung, Ingraham, Johnson-Groh, Kalas, Konopacky, Lafrenière, Larkin, Maire, Marchis, Metchev, Millar-Blanchaer, Morzinski, Nielsen, Oppenheimer, Palmer, Patel, Perrin, Poyneer, Rantakyrö, Ruffio, Savransky, Schneider, Sivaramakrishnan, Song, Soummer, Thomas, Vasisht, Wallace, Wang, Wiktorowicz, \& Wolff}]{Rajan_2017}
Rajan, A., Rameau, J., Rosa, R. J.~D., {et~al.} 2017, The Astronomical Journal, 154, 10, \dodoi{10.3847/1538-3881/aa74db}

\bibitem[{Rauscher(2024)}]{Rauscher_2024}
Rauscher, B.~J. 2024, Publications of the Astronomical Society of the Pacific, 136, 015001, \dodoi{10.1088/1538-3873/ad1b36}

\bibitem[{Rigby {et~al.}(2023)Rigby, Perrin, McElwain, Kimble, Friedman, Lallo, Doyon, Feinberg, Ferruit, Glasse, Rieke, Rieke, Wright, Willott, Colon, Milam, Neff, Stark, Valenti, Abell, Abney, Abul-Huda, Scott~Acton, Adams, Adler, Aguilar, Ahmed, Albert, Alberts, Aldridge, Allen, Altenburg, Álvarez Márquez, Alves~de Oliveira, Andersen, Anderson, Anderson, Argyriou, Armstrong, Arribas, Artigau, Arvai, Atkinson, Bacon, Bair, Banks, Barrientes, Barringer, Bartosik, Bast, Baudoz, Beatty, Bechtold, Beck, Bergeron, Bergkoetter, Bhatawdekar, Birkmann, Blazek, Blome, Boccaletti, Böker, Boia, Bonaventura, Bond, Bosley, Boucarut, Bourque, Bouwman, Bower, Bowers, Boyer, Bradley, Brady, Braun, Breda, Bresnahan, Bright, Britt, Bromenschenkel, Brooks, Brooks, Brown, Brown, Brown, Bunker, Burger, Bushouse, Cale, Cameron, Cameron, Canipe, Caplinger, Caputo, Cara, Carey, Carniani, Carrasquilla, Carruthers, Case, Catherine, Chance, Chapman, Charlot, Charlow, Chayer, Chen, Cherinka, Chichester, Chilton, Chonis, Clampin,
  Clark, Clark, Coe, Coleman, Comber, Comeau, Connolly, Cooper, Cooper, Coppock, Correnti, Cossou, Coulais, Coyle, Cracraft, Curti, Cuturic, Davis, Davis, Dean, DeLisa, deMeester, Dencheva, Dencheva, DePasquale, Deschenes, Hunor~Detre, Diaz, Dicken, DiFelice, Dillman, Dixon, Doggett, Donaldson, Douglas, DuPrie, Dupuis, Durning, Easmin, Eck, Edeani, Egami, Ehrenwinkler, Eisenhamer, Eisenhower, Elie, Elliott, Elliott, Ellis, Engesser, Espinoza, Etienne, Etxaluze, Falini, Feeney, Ferry, Filippazzo, Fincham, Fix, Flagey, Florian, Flynn, Fontanella, Ford, Forshay, Fox, Franz, Fu, Fullerton, Galkin, Galyer, García~Marín, Gardner, Gardner, Garland, Garrett, Gasman, Gaspar, Gaudreau, Gauthier, Geers, Geithner, Gennaro, Giardino, Girard, Giuliano, Glassmire, Glauser, Glazer, Godfrey, Golimowski, Gollnitz, Gong, Gonzaga, Gordon, Gordon, Goudfrooij, Greene, Greenhouse, Grimaldi, Groebner, Grundy, Guillard, Gutman, Ha, Haderlein, Hagedorn, Hainline, Haley, Hami, Hamilton, Hammel, Hansen, Harkins, Harr, Hart, Hart,
  Hartig, Hashimoto, Haskins, Hathaway, Havey, Hayden, Hecht, Heller-Boyer, Henriques, Henry, Hermann, Hernandez, Hesman, Hicks, Hilbert, Hines, Hoffman, Holfeltz, Holler, Hoppa, Hott, Howard, Howard, Hunter, Hunter, Hurst, Husemann, Hustak, Ilinca~Ignat, Illingworth, Irish, Jackson, Jahromi, Jakobsen, James, James, Januszewski, Jenkins, Jirdeh, Johnson, Johnson, Jones, Jones, Jones, Jones, Jordan, Jordan, Jurczyk, Jurling, Kaleida, Kalmanson, Kammerer, Kang, Kao, Karakla, Kavanagh, Kelly, Kendrew, Kennedy, Kenny, Keski-kuha, Keyes, Kidwell, Kinzel, Kirk, Kirkpatrick, Kirshenblat, Klaassen, Knapp, Scott~Knight, Knollenberg, Koehler, Koekemoer, Kovacs, Kulp, Kumari, Kyprianou, La~Massa, Labador, Labiano, Lagage, Lajoie, Lallo, Lam, Lamb, Lambros, Lampenfield, Langston, Larson, Law, Lawrence, Lee, Leisenring, Lepo, Leveille, Levenson, Levine, Levy, Lewis, Lewis, Libralato, Lightsey, Link, Liu, Lo, Lockwood, Logue, Long, Long, Loomis, Lopez-Caniego, Lorenzo~Alvarez, Love-Pruitt, Lucy, Luetzgendorf, Maghami,
  Maiolino, Major, Malla, Malumuth, Manjavacas, Mannfolk, Marrione, Marston, Martel, Maschmann, Masci, Masciarelli, Maszkiewicz, Mather, McKenzie, McLean, McMaster, Melbourne, Meléndez, Menzel, Merz, Meyett, Meza, Miskey, Misselt, Moller, Morrison, Morse, Moseley, Mosier, Mountain, Mueckay, Mueller, Mullally, Murphy, Murray, Murray, Mustelier, Muzerolle, Mycroft, Myers, Myrick, Nanavati, Nance, Nayak, Naylor, Nelan, Nickson, Nielson, Nieto-Santisteban, Nikolov, Noriega-Crespo, O’Shaughnessy, O’Sullivan, Ochs, Ogle, Oleszczuk, Olmsted, Osborne, Ottens, Owens, Pacifici, Pagan, Page, Park, Parrish, Patapis, Paul, Pauly, Pavlovsky, Pedder, Peek, Pena-Guerrero, Penanen, Perez, Perna, Perriello, Phillips, Pietraszkiewicz, Pinaud, Pirzkal, Pitman, Piwowar, Platais, Player, Plesha, Pollizi, Polster, Pontoppidan, Porterfield, Proffitt, Pueyo, Pulliam, Quirt, Quispe~Neira, Ramos~Alarcon, Ramsay, Rapp, Rapp, Rauscher, Ravindranath, Rawle, Regan, Reichard, Reis, Ressler, Rest, Reynolds, Rhue, Richon, Rickman,
  Ridgaway, Ritchie, Rix, Robberto, Robinson, Robinson, Robinson, Rock, Rodriguez, Rodriguez Del~Pino, Roellig, Rohrbach, Roman, Romelfanger, Rose, Roteliuk, Roth, Rothwell, Rowlands, Roy, Royer, Royle, Rui, Rumler, Runnels, Russ, Rustamkulov, Ryden, Ryer, Sabata, Sabatke, Sabbi, Samuelson, Sapp, Sappington, Sargent, Sauer, Scheithauer, Schlawin, Schlitz, Schmitz, Schneider, Schreiber, Schulze, Schwab, Scott, Sembach, Shanahan, Shaughnessy, Shaw, Shawger, Shay, Sheehan, Shen, Sherman, Shiao, Shih, Shivaei, Sienkiewicz, Sing, Sirianni, Sivaramakrishnan, Skipper, Sloan, Slocum, Slowinski, Smith, Smith, Smith, Smith, Snyder, Soh, Tony~Sohn, Soto, Spencer, Stallcup, Stansberry, Starr, Starr, Stewart, Stiavelli, Straughn, Strickland, Stys, Summers, Sun, Sunnquist, Swade, Swam, Swaters, Swoish, Taylor, Taylor, Te~Plate, Tea, Teague, Telfer, Temim, Thatte, Thompson, Thompson, Thomson, Tikkanen, Tippet, Todd, Toolan, Tran, Trejo, Truong, Tsukamoto, Tustain, Tyra, Ubeda, Underwood, Uzzo, Van~Campen, Vandal,
  Vandenbussche, Vila, Volk, Wahlgren, Waldman, Walker, Wander, Warfield, Warner, Wasiak, Watkins, Weaver, Weilert, Weiser, Weiss, Weissman, Welty, West, Wheate, Wheatley, Wheeler, White, Whiteaker, Whitehouse, Whiteleather, Whitman, Williams, Willmer, Willoughby, Wilson, Wirth, Wislowski, Wolf, Wolfe, Wolff, Workman, Wright, Wu, Wu, Wymer, Yates, Yeager, Yeates, Yerger, Yoon, Young, Yu, Zak, Zeidler, Zhou, Zielinski, Zincke, \& Zonak}]{Rigby2023}
Rigby, J., Perrin, M., McElwain, M., {et~al.} 2023, Publications of the Astronomical Society of the Pacific, 135, 048001, \dodoi{10.1088/1538-3873/acb293}

\bibitem[{Rosa {et~al.}(2015)Rosa, Nielsen, Blunt, Graham, Konopacky, Marois, Pueyo, Rameau, Ryan, Wang, Bailey, Chontos, Fabrycky, Follette, Macintosh, Marchis, Ammons, Arriaga, Chilcote, Cotten, Doyon, Duchêne, Esposito, Fitzgerald, Gerard, Goodsell, Greenbaum, Hibon, Ingraham, Johnson-Groh, Kalas, Lafrenière, Maire, Metchev, Millar-Blanchaer, Morzinski, Oppenheimer, Patel, Patience, Perrin, Rajan, Rantakyrö, Ruffio, Schneider, Sivaramakrishnan, Song, Tran, Vasisht, Ward-Duong, \& Wolff}]{deRosa2015}
Rosa, R. J.~D., Nielsen, E.~L., Blunt, S.~C., {et~al.} 2015, The Astrophysical Journal Letters, 814, L3, \dodoi{10.1088/2041-8205/814/1/L3}

\bibitem[{Ruffio {et~al.}(2019)Ruffio, Macintosh, Konopacky, Barman, De~Rosa, Wang, Hoch, Czekala, \& Marois}]{Ruffio_2019}
Ruffio, J.-B., Macintosh, B., Konopacky, Q.~M., {et~al.} 2019, The Astronomical Journal, 158, 200, \dodoi{10.3847/1538-3881/ab4594}

\bibitem[{Ruffio {et~al.}(2021)Ruffio, Konopacky, Barman, Macintosh, Hoch, De~Rosa, Wang, Czekala, \& Marois}]{Ruffio_2021}
Ruffio, J.-B., Konopacky, Q.~M., Barman, T., {et~al.} 2021, The Astronomical Journal, 162, 290, \dodoi{10.3847/1538-3881/ac273a}

\bibitem[{Ruffio {et~al.}(2024)Ruffio, Perrin, Hoch, Kammerer, Konopacky, Pueyo, Madurowicz, Rickman, Theissen, Agrawal, Greenbaum, Miles, Barman, Balmer, Llop-Sayson, Girard, Rebollido, Soummer, Allen, Anderson, Beichman, Bellini, Bryden, Espinoza, Glidden, Huang, Lewis, Libralato, Louie, Sohn, Seager, van~der Marel, Wakeford, Watkins, Ygouf, \& Mountain}]{Ruffio_2024}
Ruffio, J.-B., Perrin, M.~D., Hoch, K. K.~W., {et~al.} 2024, The Astronomical Journal, 168, 73, \dodoi{10.3847/1538-3881/ad5281}

\bibitem[{Rustamkulov {et~al.}(2022)Rustamkulov, Sing, Liu, \& Wang}]{Rustamkulov_2022}
Rustamkulov, Z., Sing, D.~K., Liu, R., \& Wang, A. 2022, The Astrophysical Journal Letters, 928, L7, \dodoi{10.3847/2041-8213/ac5b6f}

\bibitem[{{Samland, M.} {et~al.}(2017){Samland, M.}, {Mollière, P.}, {Bonnefoy, M.}, {Maire, A.-L.}, {Cantalloube, F.}, {Cheetham, A. C.}, {Mesa, D.}, {Gratton, R.}, {Biller, B. A.}, {Wahhaj, Z.}, {Bouwman, J.}, {Brandner, W.}, {Melnick, D.}, {Carson, J.}, {Janson, M.}, {Henning, T.}, {Homeier, D.}, {Mordasini, C.}, {Langlois, M.}, {Quanz, S. P.}, {van Boekel, R.}, {Zurlo, A.}, {Schlieder, J. E.}, {Avenhaus, H.}, {Beuzit, J.-L.}, {Boccaletti, A.}, {Bonavita, M.}, {Chauvin, G.}, {Claudi, R.}, {Cudel, M.}, {Desidera, S.}, {Feldt, M.}, {Fusco, T.}, {Galicher, R.}, {Kopytova, T. G.}, {Lagrange, A.-M.}, {Le Coroller, H.}, {Martinez, P.}, {Moeller-Nilsson, O.}, {Mouillet, D.}, {Mugnier, L. M.}, {Perrot, C.}, {Sevin, A.}, {Sissa, E.}, {Vigan, A.}, \& {Weber, L.}}]{Samland2017}
{Samland, M.}, {Mollière, P.}, {Bonnefoy, M.}, {et~al.} 2017, A\&A, 603, A57, \dodoi{10.1051/0004-6361/201629767}

\bibitem[{Sarkar {et~al.}(2024)Sarkar, Madhusudhan, Constantinou, \& Holmberg}]{Sarkar2024}
Sarkar, S., Madhusudhan, N., Constantinou, S., \& Holmberg, M. 2024, Monthly Notices of the Royal Astronomical Society, 531, 2731, \dodoi{10.1093/mnras/stae1230}

\bibitem[{Scarsdale {et~al.}(2024)Scarsdale, Wogan, Wakeford, Wallack, Batalha, Alderson, Aguichine, Wolfgang, Teske, Moran, López-Morales, Kirk, Gordon, Gao, Batalha, Alam, \& Adams~Redai}]{Scarsdale_2024}
Scarsdale, N., Wogan, N., Wakeford, H.~R., {et~al.} 2024, The Astronomical Journal, 168, 276, \dodoi{10.3847/1538-3881/ad73cf}

\bibitem[{Schlawin {et~al.}(2020)Schlawin, Leisenring, Misselt, Greene, McElwain, Beatty, \& Rieke}]{Schlawin_2020}
Schlawin, E., Leisenring, J., Misselt, K., {et~al.} 2020, The Astronomical Journal, 160, 231, \dodoi{10.3847/1538-3881/abb811}

\bibitem[{Schmidt {et~al.}(2025)Schmidt, MacDonald, Tsai, Radica, Wang, Ahrer, Bell, Fisher, Thorngren, Wogan, May, Ferrari, Bennett, Rustamkulov, López-Morales, \& Sing}]{schmidt2025}
Schmidt, S.~P., MacDonald, R.~J., Tsai, S.-M., {et~al.} 2025.
\newblock \doarXiv{2501.18477}

\bibitem[{Sikora {et~al.}(2024)Sikora, Rowe, Splinter, Barat, Dang, Cowan, Barclay, Colón, Désert, Kane, Llama, Shivkumar, Stassun, \& Quintana}]{sikora2024}
Sikora, J.~T., Rowe, J.~F., Splinter, J., {et~al.} 2024, Seasonal Changes in the Atmosphere of HD 80606b Observed with JWST's NIRSpec/G395H.
\newblock \doarXiv{2407.12456}

\bibitem[{Simon \& Schaefer(2011)}]{Simon2011}
Simon, M., \& Schaefer, G.~H. 2011, The Astrophysical Journal, 743, 158, \dodoi{10.1088/0004-637X/743/2/158}

\bibitem[{Snellen(2025)}]{snellen2025}
Snellen, I. 2025.
\newblock \doarXiv{2505.08926}

\bibitem[{Snellen {et~al.}(2014)Snellen, Brandl, de~Kok, Brogi, Birkby, \& Schwarz}]{Snellen_2014}
Snellen, I. A.~G., Brandl, B.~R., de~Kok, R.~J., {et~al.} 2014, Nature, 509, 63–65, \dodoi{10.1038/nature13253}

\bibitem[{{Stolker} {et~al.}(2020){Stolker}, {Quanz}, {Todorov}, {K{\"u}hn}, {Molli{\`e}re}, {Meyer}, {Currie}, {Daemgen}, \& {Lavie}}]{Stolker2020}
{Stolker}, T., {Quanz}, S.~P., {Todorov}, K.~O., {et~al.} 2020, \aap, 635, A182, \dodoi{10.1051/0004-6361/201937159}

\bibitem[{Virtanen {et~al.}(2020)Virtanen, Gommers, Oliphant, Haberland, Reddy, Cournapeau, Burovski, Peterson, Weckesser, Bright, {van der Walt}, Brett, Wilson, Millman, Mayorov, Nelson, Jones, Kern, Larson, Carey, Polat, Feng, Moore, {VanderPlas}, Laxalde, Perktold, Cimrman, Henriksen, Quintero, Harris, Archibald, Ribeiro, Pedregosa, {van Mulbregt}, \& {SciPy 1.0 Contributors}}]{2020SciPy}
Virtanen, P., Gommers, R., Oliphant, T.~E., {et~al.} 2020, Nature Methods, 17, 261, \dodoi{10.1038/s41592-019-0686-2}

\bibitem[{Wallack {et~al.}(2024)Wallack, Batalha, Alderson, Scarsdale, Adams~Redai, Aguichine, Alam, Gao, Wolfgang, Batalha, Kirk, López-Morales, Moran, Teske, Wakeford, \& Wogan}]{Wallack_2024}
Wallack, N.~L., Batalha, N.~E., Alderson, L., {et~al.} 2024, The Astronomical Journal, 168, 77, \dodoi{10.3847/1538-3881/ad3917}

\bibitem[{{Wang} {et~al.}(2021){Wang}, {Kulikauskas}, \& {Blunt}}]{whereistheplanet}
{Wang}, J.~J., {Kulikauskas}, M., \& {Blunt}, S. 2021.
\newblock \url{https://ui.adsabs.harvard.edu/abs/2021ascl.soft01003W}

\bibitem[{Wang {et~al.}(2021)Wang, Ruffio, Morris, Delorme, Jovanovic, Pezzato, Echeverri, Finnerty, Hood, Zanazzi, Bryan, Bond, Cetre, Martin, Mawet, Skemer, Baker, Xuan, Wallace, Wang, Bartos, Blake, Boden, Buzard, Calvin, Chun, Doppmann, Dupuy, Duchêne, Feng, Fitzgerald, Fortney, Freedman, Knutson, Konopacky, Lilley, Liu, Lopez, Lupu, Marley, Meshkat, Miles, Millar-Blanchaer, Ragland, Roy, Ruane, Sappey, Schofield, Weiss, Wetherell, Wizinowich, \& Ygouf}]{Wang_2021}
Wang, J.~J., Ruffio, J.-B., Morris, E., {et~al.} 2021, The Astronomical Journal, 162, 148, \dodoi{10.3847/1538-3881/ac1349}

\bibitem[{Whiteford {et~al.}(2023)Whiteford, Glasse, Chubb, Kitzmann, Ray, Phillips, Biller, Palmer, Rice, Waldmann, Changeat, Skaf, Wang, Edwards, \& Al-Refaie}]{Whiteford2023}
Whiteford, N., Glasse, A., Chubb, K.~L., {et~al.} 2023, Monthly Notices of the Royal Astronomical Society, 525, 1375, \dodoi{10.1093/mnras/stad670}

\end{thebibliography}
\bibliographystyle{aasjournal}

\end{document}